\documentclass[amsmath,amssymb,aps,reprint,floatfix,superscriptaddress]{revtex4-2}
\usepackage{amsmath,amsfonts,amssymb}
\usepackage{mathtools}
\usepackage[colorlinks=true,citecolor=blue,linkcolor=blue]{hyperref}
\usepackage{bbm}
\usepackage{braket}
\usepackage{physics}
\usepackage{comment}
\usepackage{fancyref}
\usepackage{upgreek}
\usepackage{esint}
\usepackage{soul}
\usepackage{nicefrac}
\usepackage[normalem]{ulem} 
\usepackage{floatrow}
\usepackage{graphicx}
\usepackage{dcolumn}
\usepackage{bm}
\usepackage[none]{hyphenat}
\usepackage[mathlines]{lineno}
\usepackage [english]{babel}
\usepackage{comment}
\usepackage [autostyle, english = american]{csquotes}
\MakeOuterQuote{"}

\usepackage{xcolor}
\usepackage{xspace}

\definecolor{lightblue}{cmyk}{1,0.3,0,0.2}

\usepackage[nolist]{acronym}

\newcommand{\leri}[1]{\left(#1 \right)}
\newcommand{\obs}[1]{\langle #1 \rangle }

\begin{document}

\title{Extending the Mpemba effect to the underdamped realm}

\author{Shahaf Aharony Shapira}
\email{shahaf.aharony@weizmann.ac.il}
\affiliation{Department of Physics of Complex Systems, Weizmann Institute of Science, Rehovot 7610001, Israel}
\author{Gene Chen}
\email{wkp5pp@virginia.edu}
\affiliation{Department of Physics, University of Virginia, 382 McCormick Rd, Charlottesville, 22904, VA, USA}
\author{Marija Vucelja}
\email{mvucelja@virginia.edu}
\affiliation{Department of Physics, University of Virginia, 382 McCormick Rd, Charlottesville, 22904, VA, USA}
\affiliation{Max Planck Institute for the Physics of Complex Systems, N\"othnitzer Str. 38, Dresden, 01187, Germany}
\affiliation{Department of Mathematics, University of Virginia, 141 Cabel Drive, Charlottesville, 22904, VA, USA}
\author{Oren Raz}
\email{oren.raz@weizmann.ac.il}
\affiliation{Department of Physics of Complex Systems, Weizmann Institute of Science, Rehovot 7610001, Israel}

\begin{abstract}
The Mpemba effect is the counterintuitive phenomenon in which an initially hotter system cools faster than a colder, otherwise identical system. It has been experimentally demonstrated in various classical overdamped systems. Here, we explore the existence of the same effect in a regime where inertia cannot be neglected, namely the underdamped regime. We consider the underdamped dynamics of a Brownian particle in a potential. We show perturbatively that, if the effect exists in the overdamped limit, it persists for sufficiently large but finite damping. In the ultra-weak-damping limit, we show that the effect cannot occur for smooth confining single-well potentials with canonical initial states,
but can arise in more complex potentials. We demonstrate our results numerically using double-well potentials, the canonical setting for the Mpemba effect in the overdamped limit.
\end{abstract}

\maketitle

\onecolumngrid


\begin{acronym}
    \acro{ME}{Mpemba effect}
    \acro{DoF}{degree of freedom}
    \acro{NTE}{negative thermal expansion}
    \acro{KK}{Klein-Kramers}
    \acro{SM}{Supplemental Material}
    \acro{PDF}{probability density function}
    \acro{MCF}{Matrix Continued Fractions}
    \acro{1D}{one-dimensional}
    \acro{2D}{two-dimensional}
    \acro{PDE}{partial derivative equation}
\end{acronym}



\section{Introduction}
Common sense dictates that a system initiated at a temperature closer to that of a heat reservoir will equilibrate faster than a system initiated at a farther temperature. However, as realized already in ancient times~\cite{Aristotle1989Meteorologica}, this is not always true, although for a very long time it had only been observed in water. The case in which, under certain conditions, an initially hotter system cools down faster than a colder, otherwise identical system is known as the \ac{ME}~\cite{Mpemba1969cool}. Recently, the \ac{ME} has been the subject of many studies \cite{Lasanta2017When,Lu2017Significance,klich2018solution,baity2019mpemba,Kumar2020Exponentially,Kumar2022Anomalous,Carollo2021Exponentially,Nava2019Lindblad,Murciano2023Entanglement,Walker_2021,holtzman2022landau,gal2020precooling,degunther2022anomalous,busiello2021inducing,PhysRevLett.129.138002,zhang2022theoretical,PhysRevLett.131.017101,blom2026strong,biswas2020mpemba,santos2020mpemba,chetrite2021metastable,PhysRevLett.133.010403,biswas2023mpemba,TEZA20261,biswas2023mpembaEffect,avitan2026necessary}, a review~\cite{TEZA20261}, and a unifying framework in terms of resource theory~\cite{summer_resource_2025}. Additionally, several quantum analogs have been proposed, e.g.~\cite{PhysRevLett.133.010403,zhang2025observation,joshi2024observing}, and reviewed~\cite{ares2025quantum}.

In this manuscript, we focus on classical Markovian systems, where the effect is characterized in the long-time limit, namely as a final-relaxation \ac{ME}. In such settings, the \ac{ME} has been demonstrated across a broad class of models, ranging from Markov jump processes~\cite{Klich2019Mpemba,bera2026effect,walker2023optimal,klich2018solution} to continuous-state Markov processes. In particular, the effect has been extensively studied in overdamped \ac{1D} systems. Experimentally, it has been observed in colloidal particles diffusing in external potentials~\cite{Kumar2020Exponentially,Kumar2022Anomalous}. Theoretically, it has been analyzed for Langevin particles evolving in both double-well~\cite{Walker_2021,walker2022mpemba,biswas2023mpemba} and single-well potentials~\cite{biswas2023mpembaEffect}, while the distinction between these settings was studied in Refs.~\cite{liu2026mpembahit,liu2026predictingconditionsobservingmpemba}. Metastability has been shown to play an important role in \ac{ME}~\cite{Lu2017Significance,chetrite2021metastable,Kumar2020Exponentially,Kumar2022Anomalous}, although it is not necessary for the effect to occur~\cite{Walker_2021,biswas2023mpembaEffect,liu2026mpembahit}. Recently, \ac{ME} with quantum tunneling in a double-well has been considered~\cite{melles2026quantizationclassicalmpembaeffect,hayakawa2026quantumtunnelingmpembaeffect}. 
Numerically, the effect has been investigated using Brownian dynamics simulations of both passive~\cite{malhotra2024double} and active~\cite{PhysRevLett.129.138002,biswas2023mpembaEffect} colloidal particles, with and without resetting~\cite{busiello2021inducing}, as well as in systems where coupling to the thermal bath occurs only through the system boundaries~\cite{PhysRevLett.131.017101}. More recently, the \ac{ME} has also been demonstrated in Langevin dynamics within a confined \ac{2D} potential~\cite{hayakawa2026mpemba}.

Although in recent years significant progress has been made in understanding the Mpemba effect as an abstract concept, much of this progress is not known to be directly relevant to the specific case of water, which is the historical context of the effect. First, the effect in water is observed mostly by monitoring the phase transition time, not the final relaxation time, and so far, very little effort has been devoted to the Mpemba effect through phase transitions \cite{holtzman2022landau,yang2022mpemba,yang2020non,li2026minimal}.  Moreover, it is not known what the relations are between final-time relaxation \ac{ME} (as studied in Markovian dynamics) and the effects through a phase transition. On the theoretical side, \ac{ME} in water has been numerically observed in molecular dynamics simulations of relatively small molecular systems and short time scales~\cite{jin2015mechanisms,tao2017different,ghosh2025simulations}. However, these numerical results provide little insight into the specific mechanism behind the Mpemba effect in water. 

Exploring the Mpemba effect in water using the Markovian framework is extremely difficult, since it is inherently an out-of-equilibrium many-body problem. A common approach in such cases is to study a simple system that shares some basic features with the complicated system. Specifically, we use a \ac{1D} biased double-well potential, which is commonly used to  describe the effective forces exerted on the hydrogen atom by covalent and hydrogen bonds~\cite{rekik2010influence,zhang2014hydrogen}. The same potential structure often serves as a canonical example of the Mpemba effect in the overdamped case~\cite{Lu2017Significance}.  
It is also worth noting that a biased one-dimensional double-well potential can serve as a toy model for systems with negative thermal expansion (NTE)~\cite{yoshimura1991negative}, a phenomenon in which some materials, including water over an appropriate temperature range, contract rather than expand upon heating. However, the simple models used to describe NTE and those used to study the Mpemba effect in a double-well potential differ in an important respect. The Mpemba effect is invariant under spatial reflection, provided that the full dynamical problem---including the domain and boundary conditions---is reflected accordingly. Thus, if a potential $V(x)$ with given boundary conditions exhibits a Mpemba effect, then the reflected problem with potential $V(-x)$ and reflected boundary conditions exhibits the same effect, since the relaxation dynamics is mapped exactly by coordinate inversion $x\mapsto -x$ (and momentum inversion $p\mapsto -p$ in the underdamped case). By contrast, in simple NTE models the sign of the thermal expansion is tied to the temperature dependence of a physical displacement. In this sense, NTE is not invariant under reflection in the same way.
 
To date, the Mpemba effect in double-well potentials has been studied primarily in the overdamped limit. If a final-relaxation Mpemba effect in water is ultimately associated with an effective double-well structure in the molecular interaction landscape, then an underdamped description is more appropriate. There are two reasons for this. First, the relevant molecular coordinates involve atoms with finite mass, so inertial effects cannot be ruled out a priori. Second, hydrogen-bond rearrangements in water occur on ultrafast, sub-picosecond-to-picosecond time scales, where the separation of time scales required for an overdamped approximation is not necessarily justified; see, e.g.~\cite{moilanen2008water}.

In this manuscript, we investigate the extent to which the Mpemba effect can persist beyond the overdamped limit. Our analysis shows that, similarly to the heating--cooling asymmetry~\cite{dieball2026thermal,lapolla2020faster}, the \ac{ME} survives the inclusion of inertial effects. To determine the conditions under which the \ac{ME} occurs, we numerically analyze the spectrum and eigenfunctions of the \ac{KK} operator. This analysis yields a rich diagram of ``Mpemba phases'' for the system, shown in Fig.~\ref{fig:numericalKK_phaseSpace}, identifying the type of effect present at different temperatures and dissipation rates. We further demonstrate the \ac{ME} through numerical simulations of the corresponding stochastic process. Finally, we show that the \ac{ME} can persist even in the ultra-weak-damping limit, although not in single-well potentials.

The manuscript is organized as follows. In Section~\ref{sec:Underdamped framework}, we construct the mathematical underdamped framework, including stochastic and deterministic formalisms. Additionally, we discuss the asymmetric double-well potential used for our numerical results. Then, in Section~\ref{sec:spectralDecomposition}, we explain how the \ac{ME} is realized using the \ac{KK} equation, and present the `Mpemba phases' diagram from numerical decomposition of the corresponding operator. We further investigate two limiting cases of strong and weak dissipation rates, using perturbation theory and Langevin simulations: 
In the strong-dissipation limit, presented in Section~\ref{sec:strongDissipation}, the underdamped overlap coefficient is perturbatively close to its overdamped counterpart, implying that robust overdamped Mpemba effects persist for sufficiently large damping; In the weak dissipation limit, presented in Section~\ref{sec:weakDissipation}, we show that the \ac{ME} is absent for a single-well potential, but exists for more complicated potentials, such as double-well potentials. 

Lastly, our conclusions and outlook are summarized in Section~\ref{sec:conclusions}.

\section{Underdamped framework}\label{sec:Underdamped framework} 

We consider an underdamped particle diffusing in a 1D spatial potential $V(x)$ and subject to thermal noise, $\eta(t)$. The particle is coupled to a Langevin bath consisting of a friction term proportional to the velocity and white noise. The coordinate $x$ and momentum $p = m \dot{x}$ evolve according to Langevin dynamics
\begin{align}            \label{eq:Langevin}
    m \frac{\dd^2 x}{\dd t^2} &= -\gamma \frac{\dd x}{\dd t} - \frac{\partial V}{\partial x} + \sqrt{2D(T_b)} \eta (t)\,,
\end{align}    
where $m$ is the mass of the particle, $\gamma$ is the damping constant, 
$T_b$ is the bath temperature, $k_B$ is the Boltzmann constant, and $D(T_b)\equiv\gamma k_BT_b$ is the diffusion coefficient. 
The thermal noise is taken to obey Gaussian statistics, with mean and variance
\begin{align}
    \langle \eta (t)\rangle  = 0, \quad \langle \eta (t) \eta (t')\rangle = \delta (t-t').  
\end{align}
For a Brownian particle in a solvent, the noise comes from random kicks by the solvent molecules, which can often be assumed (as here) to be instantaneous and uncorrelated. 

The evolution of the \ac{PDF}, $f(x,p,t)$, is described by the~\ac{KK} equation
\begin{subequations}
\begin{align}
    \label{eqn:KleinKramers}
	\partial_t f  
    &= \mathcal{L} f\,,
    \\  
    \mathcal{L}&\equiv-\frac{p}{m} \partial_x  + \partial_p \left(V' + \frac{\gamma}{m}  p\right)+ \gamma k_B T_b \partial_p^2  \,,
\end{align}
\end{subequations}
where $V' \equiv \dd V /\dd x$ and $\mathcal{L}$ is the \ac{KK} operator. The KK equation can be viewed as a sum of Liouville equation 
and dissipative dynamics, as follows:
\begin{subequations}
\begin{align} \label{eq:kkWithEpsilon}
    \partial_t f  
    &= \acomm{\mathcal{H}}{f} + \epsilon \mathcal{H}_1 f\,, 
    \\
    \mathcal{H} (x,p) &= \frac{p^2}{2m} + V(x)\,,\\
    \mathcal{H}_1 &\equiv \omega_0 \leri{\partial_p p + m k_B T_b \partial_p^2}\,,
\end{align}
\end{subequations}
where $\{\,,\}$ are the Poisson brackets, $\mathcal{H}$ is the Hamiltonian of the system
and
$\mathcal{H}_1$ describes the coupling to the Markovian heat bath with a dimensionless coupling strength 
\begin{align} \label{eqn:epsilonDef}
    \epsilon &\equiv \frac{\gamma}{m\omega_0}\,.
\end{align}
Accordingly, $\omega_0$ has units of frequency, and in principle can be arbitrarily chosen. A natural choice is a characteristic frequency associated with the potential. There are several such frequencies in a double-well potential: In the low-energy regime, one may Taylor-expand the potential around its global minimum at $x=x_g$, yielding
$\omega_0 = \sqrt{\frac{V''(x_g)}{m}}$.
In the high-energy regime, $\omega_0$ can instead be defined through the inverse oscillation period between the classical turning points $x_1$ and $x_2$, i.e., 
$\omega_0 = \pi \left(\int_{x_1}^{x_2} \frac{\mathrm{d}x}{\dot{x}} \right)^{-1}$. In any case, the results are independent of the specific choice. 

To preserve probability, one imposes reflective boundary conditions in a finite domain or natural boundary conditions in an infinite domain. In the long time limit, the system reaches thermal equilibrium, characterized by the Boltzmann distribution: 
\begin{align}
    \label{eqn:Boltzmann-PDF}
    f_{\mathrm{eq}} (x,p,T_b) = \frac{1}{\mathcal{Z}(T_b)}e^{-\frac{\mathcal{H}(x,p)}{k_BT_b}},
\end{align}
where $\mathcal{Z}(T_b)$ is the partition function
\begin{align}
    \mathcal{Z}(T_b) &= \int_
    {\mathcal{D}_x}\dd x \int_{-\infty}^{\infty} 
      \dd p\,
     e^{-\frac{\mathcal{H}(x,p)}{k_BT_b}}\,. 
\end{align}
The spatial domain, $
\mathcal{D}_x$, can be finite or infinite.
The family of Boltzmann distributions $f_\text{eq}(x,p,T_b)$, parametrized by the bath temperature $T_b$, can be viewed as a 1D curve in the set of all normalizable distributions $f(x,p)$. We refer to this curve as the `equilibrium locus'. Each point on the equilibrium locus describes the long-time limit of a system coupled to a bath at a different temperature. 

\begin{figure}
	\centering
	\includegraphics[width=0.6\textwidth]
    {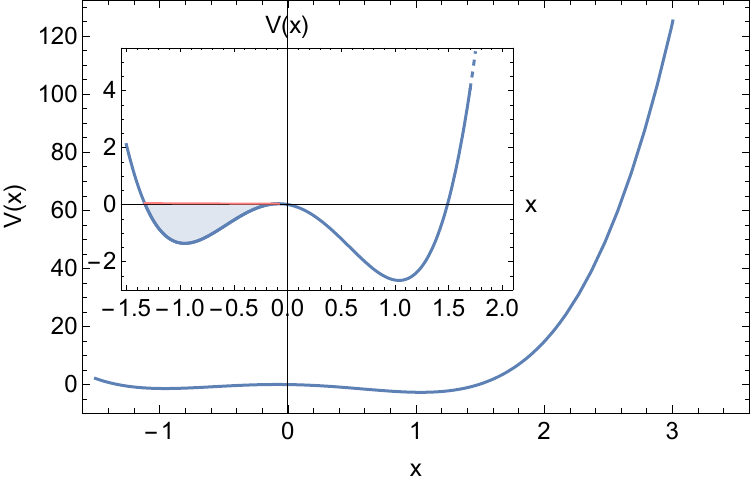}
	\caption{The potential $V(x)$ used for numerical simulations throughout this manuscript, given by Eq.~\eqref{eqn:Potential} with parameters: $d_1=-0.65,\ d_2=-8,\ d_4=8$, and the rest are zero~\cite{Kumar2020Exponentially}. We used an asymmetric spatial domain $x \in [-1.5, 3.5]$. Inset: A focus on low energies, where the structure of the asymmetric double-well is more apparent. A particle is considered to be in the left well (shaded region) for $x_\text{left}\in [-1.33, -0.08]$.}
	\label{fig:potential}
\end{figure}

As a concrete example, in the following numerical simulations, we use an asymmetric double-well potential of the form (see Fig.~\ref{fig:potential}):
\begin{align}
   \label{eqn:Potential}
    V(x) =  
    d_1 x + \frac{d_2}{2} x^2 + \frac{d_3}{3} x^3 + \frac{d_4}{4} x^4
    \,,
\end{align} 
in a spatially finite domain $x\in [x_\mathrm{min}, x_\mathrm {max}]$. In our numerical examples, we used the following values: $d_1 = -0.65$, $d_2=-8$, $d_3=0$, $d_4 = 8$, $x_\mathrm{min}=-1.5$ and $x_\mathrm{max}=3.5$. The motivation for working with this specific potential is twofold: First, potentials describing a metastable state are among the simplest known to give rise to the \ac{ME} in the overdamped regime~\cite{Kumar2020Exponentially, Kumar2022Anomalous}. It is therefore natural to use this potential when exploring whether the \ac{ME} persists in the underdamped case. Moreover, as discussed in the introduction, this potential resembles the effective potentials of hydrogen and covalent bonds~\cite{rekik2010influence} and is also commonly used in simple models of \ac{NTE}.

\section{The Mpemba effect through spectral decomposition of the {KK} equation}\label{sec:spectralDecomposition}

\begin{figure}[t]
	\centering
	\includegraphics[width=0.6\textwidth]
    {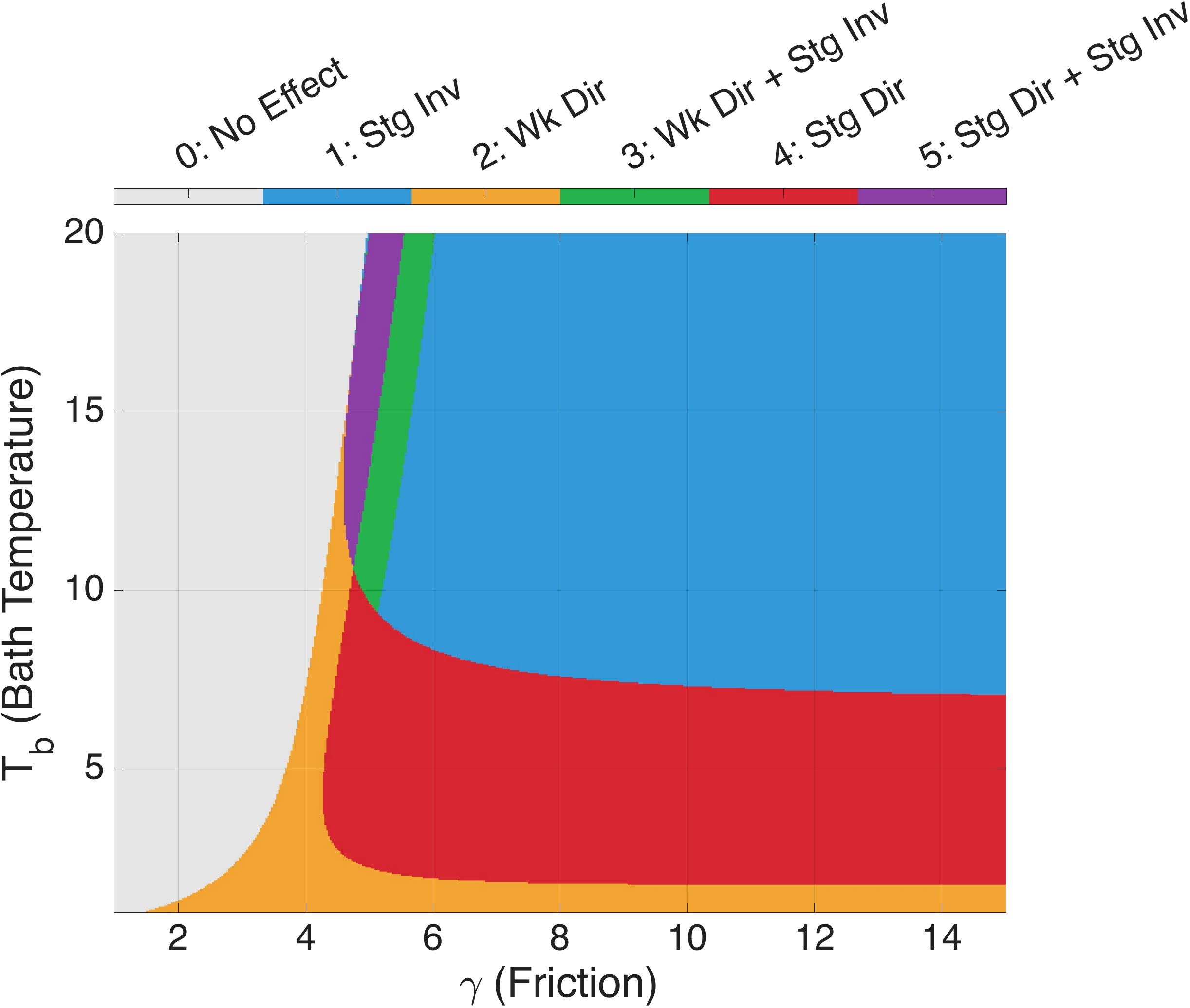}
	\caption{The `Mpemba phases' diagram of an underdamped asymmetric double-well potential. We present the type of \ac{ME}s (color) for every set of dissipation $\gamma$ (x-axis) and bath temperature $T_b$ (y-axis). Given these parameters, the \ac{KK} operator was discretized, diagonalized, and the coefficient of slow relaxation $a_2$ was evaluated. When the leading nonzero eigenvalue is complex, the corresponding norm of the projection onto the leading 2D eigenspace is used instead. Nonmonotonicity (vanishing) indicates a weak (strong) effect, and a cooling (heating) is referred to as a direct (inverse) effect. Details of the numerics are given in~\cite{SM}.
    Relative to the potential (Fig.~\ref{fig:potential}), weak (strong) dissipation is considered for $\gamma\ll \leri{\gg} 3$.}
	\label{fig:numericalKK_phaseSpace}
\end{figure}

The \ac{KK} equation can be solved by spectral decomposition of the \ac{KK} operator $\mathcal{L}$, with 
\begin{align}
    \label{eqn:KK-eigenvalue-problem}
    \mathcal{L}\, v_n(x,p,T_b) = \lambda_n(T_b) v_n(x,p,T_b), 
\end{align}
where $v_n$ are the right eigenfunctions of the \ac{KK} operator, and $\lambda_n$ are their corresponding eigenvalues. 
The eigenvalues are ordered with $\lambda_1 =0 > \mathrm{Re}(\lambda_2) \geq \mathrm{Re}(\lambda_3) \geq \dots$
The first eigenvector, $v_1$, is proportional to the Boltzmann distribution $f_{\mathrm{eq}} (x,p,T_b)$. The operator is not Hermitian, thus having different left eigenfunctions defined through
\begin{align}
     \mathcal{L}^\dagger u_n(x,p,T_b) = \lambda_n(T_b) u_n(x,p,T_b)\,,
\end{align}
where $\mathcal{L}^\dagger$ is the adjoint of the ~\ac{KK} operator. 
The right and left eigenfunctions are related through
\begin{align}
    \label{eq:left-rigth-eigenfunctions}
    u_n (x,p,T_b) = \mathcal{Z}(T_b) 
    e^{\frac{\mathcal{H}(x,p)}{k_BT_b}} v_n(x,-p,T_b),
\end{align}
e.g.~\cite{risken1996fokker}. It is convenient to normalize the left and right eigenfunctions such that they form a biorthogonal basis. 

The PDF $f(x,p,t)$ can be written as the spectral decomposition of eigenfunctions of the \ac{KK} operator as  
\begin{align}
\label{eq:f-eigen-solution}
    f(t) = f_{\mathrm{eq}} (T_b) + \sum ^\infty _{n = 2} a_n v_n(T_b)e^{\lambda_n(T_b) t}, 
\end{align}
where $a_n$ is the projection of the initial state on the left eigenfunction $u_n(T_b)$. That is, 
\begin{align}
    a_n = \frac{\langle u_n(T_b), f_\mathrm{init}\rangle}{\langle u_n(T_b), v_n(T_b) \rangle}, 
\end{align}
where $f_\mathrm{init}$ is the initial condition and $\langle \,, \rangle$ denotes the scalar product, defined as 
\begin{align}
    \langle u,v \rangle \equiv \int _{\mathcal{D}_x}\dd x \int ^\infty _{-\infty}  \dd p\, u(x,p) v(x,p). 
\end{align}
Below, we focus on the initial case being a Boltzmann distribution at temperature $T_i$, then the coefficients $a_n$ are given by 
\begin{align} 
\label{eqn:a2Formula}
    a_n(T_i,T_b) = \frac{\langle u_n (T_b), f_{\mathrm{eq}} (T_i)\rangle}{\langle u_n(T_b),v_n(T_b)\rangle}. 
\end{align}

When the leading nonzero eigenvalue is real and nondegenerate, with
\(\mathrm{Re}\,\lambda_2>\mathrm{Re}\,\lambda_3\), the long-time relaxation is governed by a single mode,
\begin{align}
f(t)\simeq f_{\rm eq}(T_b)+a_2(T_i,T_b)v_2(T_b)e^{\lambda_2(T_b)t}.
\end{align}
The deviation from equilibrium is proportional to the overlap coefficient with the slowest relaxation mode $a_2$. A necessary and sufficient condition for the final relaxation \ac{ME} is a nonmonotonic dependence of $a_2$ on the initial temperature, for example, in cooling for 
\begin{align}
\label{eq:Mpemba-criterion}
    |a_2(T_h,T_b)| < |a_2(T_c,T_b)| \quad \text{for}
    \quad T_b < T_c < T_h \,. 
\end{align}
The criterion, Eq.~\eqref{eq:Mpemba-criterion}, for the \ac{ME} was introduced in~\cite{Lu2017Significance}. The \ac{ME} in cooling is often referred to as the direct \ac{ME}, while in heating it is called the inverse \ac{ME}~\cite{Lu2017Significance,Kumar2022Anomalous}. The effect is strongest when the overlap coefficient $a_2$ vanishes at $T_i \neq T_b$. For this initial temperature, the relaxation rate is determined by $\lambda_3$. We call this the strong \ac{ME}. The strong \ac{ME} was introduced in~\cite{Klich2019Mpemba} and first observed in~\cite{Kumar2020Exponentially}. 

Note that in parameter regimes where the leading nonzero eigenvalues form a complex-conjugate pair, the Mpemba criterion is formulated in terms of the norm of the projection onto the corresponding leading two-dimensional eigenspace. In this case, the leading deviation from equilibrium is set by the amplitude multiplying the term that decays as $\exp(-\mathrm{Re}(\lambda_2)t)$.

\begin{figure}
	\centering
	\includegraphics[width=0.6\textwidth]
    {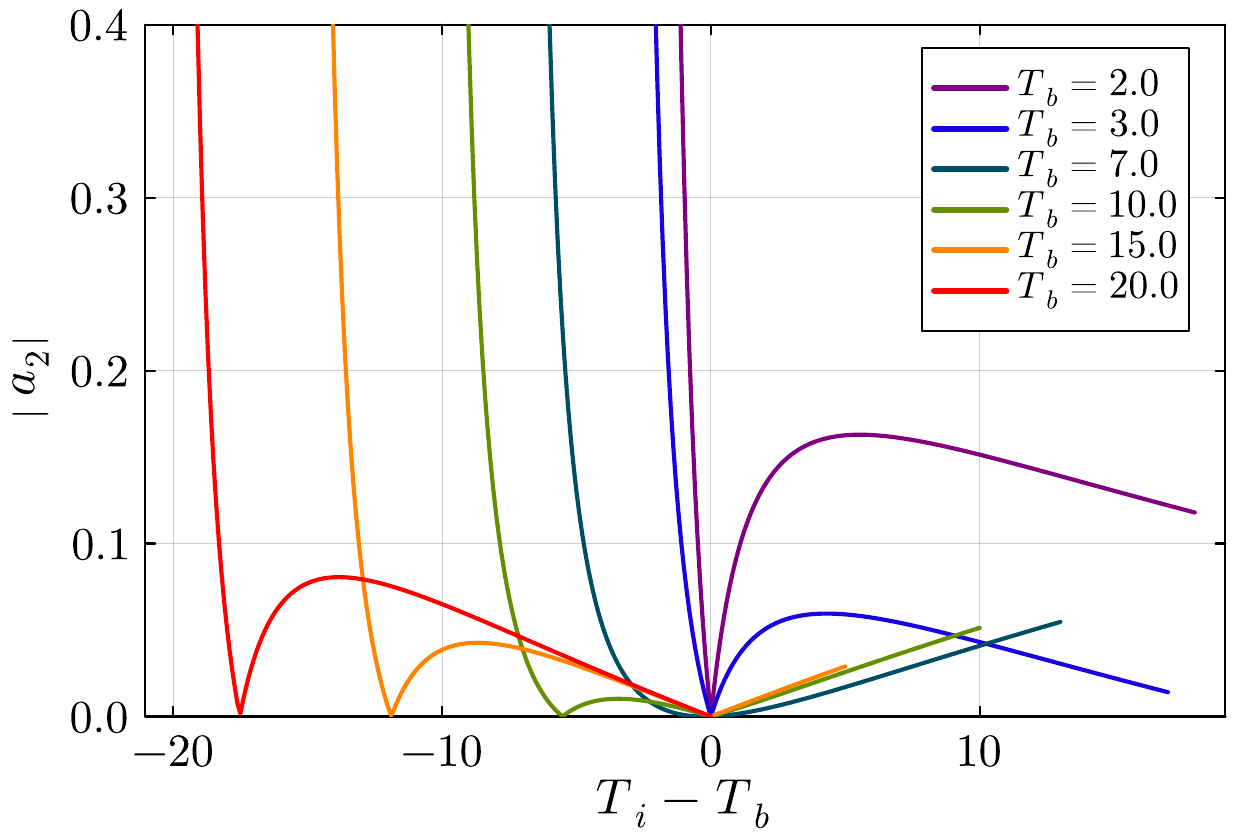}
	\caption{
    The normalized coefficient associated with the slow relaxation mode, $a_2$, in the overdamped regime as a function of the initial temperature, shown for different bath temperatures. The figure illustrates the transition between the direct ($T_b<7$) and inverse ($T_b>7$) Mpemba effects. The discretization scheme used for the corresponding Fokker--Planck operator follows~\cite{grima2004accurate}.
    }
	\label{fig:directToInverse_overdamped}
\end{figure}

Fig.~\ref{fig:numericalKK_phaseSpace} shows the `Mpemba phases' diagram, namely what type of Mpemba effects exist for each value of $T_b$ and $\gamma$. The strong dissipation limit coincides with the overdamped limit, as expected. We can also observe that for any $T_b>0$, there is some critical value of $\gamma$ below which no effect exists. At high values of $\gamma$, the transition between the weak to strong direct effect at low bath temperatures as well as the transitions from a strong direct to a strong inverse effect at around $T=7$ (see Fig.~\ref{fig:directToInverse_overdamped}) is consistent with the overdamped limit. Somewhat unexpectedly, however, we observe that lowering $\gamma$ changes the type of \ac{ME}, for example, from strong direct to strong inverse, at a constant temperature. 

Fig.~\ref{fig:numericalKK_phaseSpace} shows that, for the chosen potential, the coexistence of the direct and inverse Mpemba effects (green and purple regions) occurs only in the underdamped regime and at sufficiently high temperatures. Finally, varying the dissipation rate while keeping the bath temperature fixed can qualitatively alter the nature of the Mpemba effect. For instance, by increasing the damping rate while keeping the bath temperature constant, the system may transition from a weak direct Mpemba effect to a strong direct Mpemba effect, i.e., for example, going from orange to a red region in~Fig.~\ref{fig:numericalKK_phaseSpace}. In this case, the relaxation dynamic can change from a weak to a strong effect, and the direction from which equilibrium is approached is altered solely through the increase in damping.

\section{Strong dissipation limit} \label{sec:strongDissipation}

In the strong dissipation limit of the \ac{KK} equation~\eqref{eq:kkWithEpsilon}, momentum relaxes on a faster timescale, $\tau_p\propto m/\gamma$, compared to that of the coordinate, $\tau_x \propto \omega_0^{-1}$. In this case, the coupling strength~\eqref{eqn:epsilonDef} $\epsilon \gg 1$ and one can look for corrections to the eigenfunctions and the eigenvalues by orders of $1/\epsilon$, for details see e.g.~\cite{gardiner2009stochastic}. 

\subsection{Perturbative approach to the strong dissipation limit}\label{subsec:strongPerturbative}
The ``slow'' set of eigenvalues scale with $\epsilon$ as 
\begin{align}
\label{eq:eigs-eps}
\lambda_n = \frac{\mu_n}{\epsilon} + \mathcal{O}\left(\epsilon^{-3}\right). 
\end{align}
Here $\mu_n$ has units of inverse time and is $\epsilon-$independent. The corrections of order $\epsilon^{-2}$ are zero due to parity with respect to momentum. 

The right and left 
eigenfunctions of the \ac{KK} generator $\mathcal{L}$, 
can also be expanded in orders of $\epsilon^{-1}$ as 
\begin{subequations}
\begin{align}
v_n&= v^{(0)}_n + \frac{1}{\epsilon} v^{(1)}_n + \frac{1}{\epsilon^2}v_n ^{(2)} + \mathcal{O}\left(\epsilon^{-3}\right), 
\\
u_n&= u^{(0)}_n + \frac{1}{\epsilon} u^{(1)}_n + \frac{1}{\epsilon^2}u^{(2)}_n + \mathcal{O}\left(\epsilon^{-3}\right). 
\end{align}
\end{subequations}
From \ac{KK} equation, Eq.~\eqref{eqn:KK-eigenvalue-problem}, perturbative equations for the right eigenfunction $v_n$ are
\begin{subequations}
\begin{align}
\label{eq:KK-v-eps}
\mathcal{H}_1 v^{(0)}_n &= 0\,,  
\\
\left\{\mathcal{H}, v^{(0)}_n\right\} + \mathcal{H}_1 v^{(1)}_n &= 0\,,
\\
  \left\{\mathcal{H}, v^{(1)}_n\right\} +  \mathcal{H}_1 v^{(2)}_n &= \mu_n v^{(0)}_n\,, 
\end{align}
\end{subequations}
and analogously for the left eigenfunction $u_n$,  
\begin{subequations}
\begin{align}
    \mathcal{H}_1^\dagger  u^{(0)}_n &= 0\,,
    \\
    -\left\{\mathcal{H},u^{(0)}_n\right\} + \mathcal{H}_1 ^\dagger u^{(1)}_n &=0\,, 
    \\
    \label{eq:KKadj-u-eps-2}
    - \left\{ \mathcal{H}, u^{(1)}_n \right\} + \mathcal{H}_1^\dagger u^{(2)}_n &= \mu_n u^{(0)}_n. 
\end{align}
\end{subequations}
Solving Eqs.~\eqref{eq:KK-v-eps}--\eqref{eq:KKadj-u-eps-2} for $v_n$ and $u_n$ functions, and setting the normalization such that the  eigenfunctions are biorthogonal up to $\epsilon^{-3}$ order, we have 
\begin{align}
 \label{eq:v-large-eps}
    &v_n(x,p) = M(p) \left(v^{(S)}_n(x) -\frac{1}{\epsilon} \frac{p}{\omega_0m} A_n(x)
    +  \frac{1}{\epsilon^2}\frac{1}{2m^2 \omega _0^2}\left(p^2 - m k_BT_b\right)C_n(x)\right) + \mathcal{O} \left(\epsilon^{-3}\right)\,, 
\end{align}
where $M(p)$ is a Gaussian in momentum $p$
\begin{align}
    M(p) = \frac{1}{\sqrt{2 \pi m k_B T_b }} e^{- \frac{p^2}{2 m k_BT_b}}\,.
\end{align}
The functions $A_n(x)$ and $C_n(x)$ are 
\begin{subequations}
\begin{align}
 A_n(x) &\equiv \left(\frac{V'(x)}{k_B T_b}v^{(S)}_n(x) + \left(v^{(S)}_n  (x)\right)' \right)\,, 
\\
    C_n(x) &\equiv \left(\frac{V'(x)}{k_BT_b}A_n(x) + A_n'(x)\right)\,,  
\end{align}
\end{subequations}
While $v^{(S)}_n(x)$ is the right eigenfunction of the overdamped Fokker-Planck (Smoluchowski) generator 
\begin{align}
\mathcal{L}_S  &\equiv\frac{1}{\gamma} \partial _x (V'(x) + k_B T_b \partial_x)\,, 
\end{align}
with eigenvalue problem
\begin{align}
\mathcal{L}_S v^{(S)}_n &=  \lambda^{(S)}_nv^{(S)}_n\,,
\end{align}
and $\lambda^{(S)}_n$ the corresponding eigenvalue in the overdamped case
\begin{align}
    \label{eq:eig-Smo-eps}
    \lambda^{(S)}_n = \frac{\mu_n}{\epsilon}\,. 
\end{align}
The eigenvalue in the underdamped case is approximated as 
\begin{align}
    \lambda _n = 
    \lambda^{(S)}_n + \mathcal{O} \left(\epsilon^{-3}\right) = \frac{\mu_n}{\epsilon}+\mathcal{O} \left(\epsilon^{-3}\right) \,. 
\end{align}
The left eigenfunction $u_n(x,p)$ is 
\begin{align}
 \nonumber
    u_n(x,p) =&\,u ^{(S)}_n(x) + \frac{1}{\epsilon}\frac{p}{m\omega _0}\left(u ^{(S)}_n(x)\right)'
    \\
  \label{eq:u-large-eps} 
    &+ \frac{1}{\epsilon^2}\left(\frac{p^2 -m k_B T_b}{2 m^2 \omega _0^2 } \left(u ^{(S)}_n(x)\right)''- \frac{\mu_n}{\omega_0} u ^{(S)}_n(x)\right)+\mathcal{O}\left(\epsilon^{-3}\right),  
\end{align}
where $u ^{(S)}_n(x)$ is the left eigenfunction of the overdamped Fokker-Planck (Smoluchowski) generator. Details of the derivations for the eigenvectors $v_n$ and $u_n$, and the eigenvalue $\lambda_n$ are given in \ac{SM}~\cite{SM}. 

Assuming that the system initially starts at temperature $T_i$ and using the expressions for the eigenfunctions, Eqs.~\eqref{eq:v-large-eps} and~\eqref{eq:u-large-eps}, one can find the overlap coefficient $a_2$ in orders of $\epsilon$.   
\begin{align}\nonumber
a_2(T_i,T_b)
=& a_2^{(S)}(T_i,T_b)
+\\
&
\frac{1}{\epsilon^2}
\left[
\frac{k_B(T_i-T_b)}{2m\omega_0^2}
\left\langle
\left(u_2^{(S)}(x)\right)''\right\rangle _{f^{(S)} _
\mathrm{eq}(T_i)}
-
\frac{\mu_2}{\omega_0}a_2^{(S)}(T_i,T_b)
\right]
+
O(\epsilon^{-3})\,,
\end{align}
where $a_2 ^{(S)}(T_i,T_b)$ is the overlap coefficient in the overdamped case, and 
\begin{align}
   \left\langle \left(u_2^{(S)}(x)\right)''\right\rangle _{f^{(S)} _
\mathrm{eq}(T_i)}=  \int _
{\mathcal{D}_x}\dd x\,f^{(S)} _
\mathrm{eq}(x,T_i)\,\left(u_2^{(S)}(x)\right)''
\end{align}
is the expectation value of $\left(u_2^{(S)}(x)\right)''$ with respect to the overdamped equilibrium distribution 
\begin{align}
   f^{(S)} _
\mathrm{eq}(x,T_i) = \frac{1}{
\mathcal{Z}^{(S)}(T_i)
}e^{-\frac{V(x)}{k_BT_i}}\,,
\end{align}
at initial temperature $T_i$, with partition function 
\begin{align}
    \mathcal{Z}^{(S)}(T_i)= \int _{
\mathcal{D}_x} \dd x \, e^{-
\frac{V(x)}{k_BT_i}}\,.
\end{align}

We rewrite the above expression for the overlap $a_2(T_i,T_b)$, by grouping together the terms with the overlap in overdamped case, $a_2 ^{(S)}(T_i,T_b)$, as follows 
\begin{align}
\label{eq:a2-under-over}
a_2(T_i,T_b)
&=
\left(1 - \frac{1}{\epsilon^2}\frac{\mu_2}{\omega_0}\right)a_2^{(S)}(T_i,T_b)
+
\frac{1}{\epsilon^2}
\frac{k_B(T_i-T_b)}{2m\omega_0^2}
\left\langle
\left(u_2^{(S)}(x)\right)''\right\rangle _{f^{(S)} _
\mathrm{eq}(T_i)}
+
O(\epsilon^{-3})\,.
\end{align}
Recall that, $\mu _2 = \epsilon \lambda_2 ^{(S)}$, from Eq.~\eqref{eq:eig-Smo-eps}.

Based on Eq.~\eqref{eq:a2-under-over}, we conclude that the leading term in the underdamped overlap coefficient is the overdamped overlap coefficient. Therefore, for sufficiently large damping, any robust overdamped Mpemba effect persists under the inclusion of inertia. The finite-damping corrections are of order \(\epsilon^{-2}\) and generally shift the location of the zero of \(a_2\). Thus, the strong-Mpemba temperature is perturbatively close to, but not exactly identical to, its overdamped value. Importantly, this derivation does not rely on the specific form of the potential and is therefore valid for an arbitrary confining potential.

\subsection{Numerics in the strong dissipation limit}\label{subsec:strongNumerics}
\begin{figure}
	\centering
	\includegraphics[width=0.6\textwidth]
    {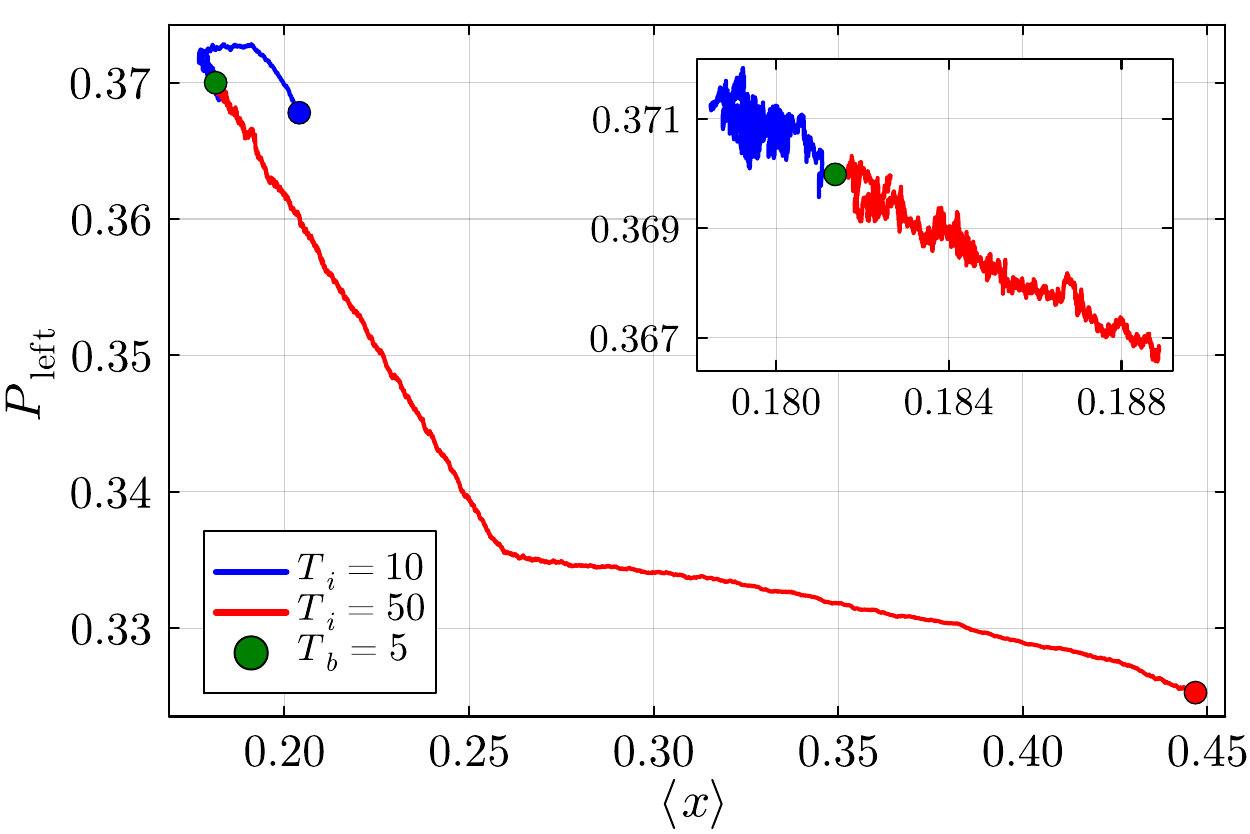}
	\caption{Cooling under strong dissipation: Relaxation from two different initial temperatures, $T_i=10$ (blue) and $T_i=50$ (red), toward the bath temperature $T_b=5$ (green dot), for a relatively large damping coefficient $\gamma=10$. For clarity, the results are projected onto the $\obs{x}$--$P_\text{left}$ plane. During the late stages of relaxation, shown in the inset, the two states approach the bath from opposite directions, 
    indicating the existence of a strong \ac{ME} at an intermediate temperature~\cite{gal2020precooling}. Further details are provided in the \ac{SM}~\cite{SM}.
    }
\label{fig:strongDissipationLangevin}
\end{figure}

Langevin simulations~\eqref{eq:Langevin} with strong dissipation agree with the perturbative computation (Section~\ref{sec:strongDissipation}) and the \ac{KK} numerics (Fig.~\ref{fig:numericalKK_phaseSpace}). We present the dynamics of a cooling process of different initial temperatures, while the remaining parameters are kept constant (see Fig.~\ref{fig:strongDissipationLangevin}). Constructing the full \ac{PDF}, a high-dimensional object, requires a multitude of numerical simulations. For simplicity and without loss of generality, we track the evolution over time of two informative, yet non-unique, parameters: the averaged position of the particle, $\obs{x} 
\equiv \iint \dd x \dd p f(x,p,t) x $, and the probability to be in the left-well, $P_\text{left}$ (see inset in Fig.~\ref{fig:potential}). We effectively project the \ac{PDF} onto a \ac{2D} plane, which also serves as an illustrative tool. Our numerical results support the conclusion from perturbation theory: In the strong dissipation limit of an underdamped system, the effect persists. 
An indication of a strong ME is obtained from the approach from opposite directions at the late stages of relaxation~\cite{gal2020precooling}.
This implies a strong \ac{ME} at an intermediate temperature between $T_i=10$ and $50$ (see Section ``Langevin dynamics'' in the \ac{SM}~\cite{SM} for more details).

\section{Weak dissipation limit}\label{sec:weakDissipation}

In the weak dissipation limit of the KK equation (Eq.~\ref{eqn:KleinKramers}), the dissipation rate is much smaller than the natural frequency of the unperturbed Hamiltonian, $\frac{\gamma}{m} \ll \omega_0$. In this regime, the system undergoes many fast Hamiltonian oscillations before exchanging a significant amount of energy with the thermal bath. Consequently, the fast intra-well oscillations can be averaged out, and the dynamics reduce to a 1D diffusion process where the slow state variable is the energy, $E$. As we argue below, this implies that for a single-well potential, there cannot be a Mpemba effect in the weak-damping limit. This result should be contrasted with the overdamped limit, where Mpemba effects are known to exist even in a single-well potential~\cite{biswas2023mpembaEffect}. However, for more complicated potentials, Mpemba effects can exist, as we demonstrate through a concrete example of a double-well potential.

\subsection{Energy diffusion}
In the weak dissipation limit, the system spreads on an energy shell much faster than the rate at which it diffuses between different energy shells. Therefore, to describe the evolution in energy space in this limit, one must account for the volume enclosed by each energy shell, which in \ac{1D} systems is given by the action, $J(E) = \oint p \, \dd x$, associated with the classical orbits. The Boltzmann distribution for a system at temperature $T$ is then given by 
\begin{align}
    F_\mathrm{eq}(E,T) = 
    \frac{1}{Z(T)}\rho(E)e^{-E/k_BT}\,,
\end{align}
where $\rho(E) = \frac{\dd J}{\dd E}$ is the density of states, and $Z(T) = \int \dd E \,\rho(E)e^{-E/k_BT}$ is the partition function.

As shown in~\cite{hanggi1990reaction}, the \ac{KK} equation in the weak damping limit can be reduced to the following Kramers energy-diffusion equation for $F(E,t)$, which is the probability of finding the system at energy $E$ at time $t$:
\begin{subequations}
\begin{align}\label{eq:KramersEnergyDiffusion}
\partial_t F(E,t) &= \frac{\gamma}{m} \partial_E \left( J(E) \left( 1 + k_B T_b \partial_E \right) \frac{F(E,t)}{\rho(E)} \right) \\
&\equiv \mathcal{L}_\mathrm{KED}F(E,t)
\end{align}
\end{subequations}
where $\mathcal{L}_\mathrm{KED}$ is the linear operator of the Kramers energy diffusion above. This equation is very similar in its structure to the overdamped Fokker-Planck equation; however, the details are somewhat different, since the `potential' in this case is not separated from the diffusion, as is the case in the overdamped Fokker-Planck equation.  

In what follows, we first discuss the single-well case and show that no type of \ac{ME} can exist for such potentials. Next, we discuss the double-well potential, explain why the same argument cannot work in this case, and present an example where there is a strong inverse Mpemba effect.

\subsection{No Mpemba effect in a single-well potential}

To establish the absence of both the strong and weak Mpemba effects in a single-well potential in the weak-dissipation limit, we analyze the nodal and monotonicity properties of the eigenfunctions governing the energy diffusion. The Kramers energy-diffusion equation, Eq.~
\eqref{eq:KramersEnergyDiffusion}, can be mapped to an eigenvalue problem. Its backward operator, $\mathcal{L}^\dagger_{\rm KED}$, which determines the left eigenfunctions $u_n(E)$, can be written in Sturm--Liouville form as
\begin{align}
\mathcal{L}^\dagger_{\rm KED}u_n
=
\frac{(\gamma/m) k_B T_b}{\rho(E)e^{-E/k_BT_b}}
\partial_E
\left[
J(E)e^{-E/k_BT_b}\partial_E u_n(E)
\right]
=
\lambda_n u_n(E).
\end{align}
It is useful to define
\begin{align}
q(E)\equiv J(E)e^{-E/k_BT_b},
\qquad
w(E)\equiv \rho(E)e^{-E/k_BT_b}.
\end{align}
For the first excited eigenfunction $u_2(E)$, whose eigenvalue satisfies $\lambda_2<0$, the eigenvalue equation can then be written as
\begin{align}
\partial_E\left[q(E)\partial_E u_2(E)\right]
=
-\kappa\, w(E)u_2(E),
\qquad
\kappa\equiv -\frac{\lambda_2}{(\gamma/m)k_BT_b}>0 .
\end{align}

By Sturm--Liouville theory, the first excited state $u_2(E)$ has exactly one zero in the interior of the energy domain. We denote this zero by $E^*$. Choosing the overall sign convention such that $u_2(E)<0$ for $E_{\min}<E<E^*$ and $u_2(E)>0$ for $E>E^*$, we now show that $u_2(E)$ is strictly increasing.

For $E<E^*$, integrating the eigenvalue equation from $E_{\min}$ to $E$ gives
\begin{align}
q(E)\partial_E u_2(E)
=
-\kappa
\int_{E_{\min}}^E dE'\, w(E')u_2(E') .
\end{align}
Here we used the natural boundary condition at the minimum-energy boundary. Since $w(E)>0$ and $u_2(E)<0$ throughout the interval $E_{\min}<E<E^*$, the integral on the right-hand side is negative. Therefore,
\begin{align}
q(E)\partial_E u_2(E)>0,
\qquad E_{\min}<E<E^* .
\end{align}
Since $q(E)>0$ in the interior of the domain, it follows that
\begin{align}
\partial_E u_2(E)>0,
\qquad E_{\min}<E<E^* .
\end{align}

Similarly, for $E>E^*$, integrating the same equation from $E$ to infinity and using the natural boundary condition at large energy gives
\begin{align}
q(E)\partial_E u_2(E)
=
\kappa
\int_E^\infty dE'\, w(E')u_2(E') .
\end{align}
In this region $u_2(E)>0$, and therefore the integral is positive. Hence
\begin{align}
q(E)\partial_E u_2(E)>0,
\qquad E>E^*,
\end{align}
and again, since $q(E)>0$ in the interior of the domain,
\begin{align}
\partial_E u_2(E)>0,
\qquad E>E^* .
\end{align}
We thus conclude that the first excited left eigenfunction $u_2(E)$ is a strictly monotonically increasing function of the energy throughout the single-well energy domain.

The overlap coefficient of the Kramers energy-diffusion equation, $\tilde a_2(T_i,T_b)$ (we used the tilde sign to differentiate it from the overlap coefficient of the KK equation), is the projection of the initial Boltzmann distribution at temperature $T_i$ onto this left eigenfunction:
\begin{align}
\tilde a_2(T_i,T_b)
=
\frac{1}{Z(T_i)}
\int_{E_{\min}}^\infty dE\,
u_2(E)\rho(E)e^{-E/k_BT_i}.
\end{align}
By the biorthogonality of the eigenfunctions, $\tilde a_2(T_b,T_b)=0$. Since $u_2(E)$ is strictly increasing and has a single zero, increasing $T_i$ shifts statistical weight toward larger energies and therefore increases the value of the overlap. Thus $\tilde a_2(T_i,T_b)$ can cross zero only at $T_i=T_b$, excluding a strong Mpemba effect in a single-well potential in the weak-dissipation limit.

The same monotonicity also excludes the weak Mpemba effect. To see this explicitly, write the derivative of $a_2$ with respect to the inverse temperature $\beta_i\equiv 1/k_BT_i$ as
\begin{align}
\frac{\partial \tilde a_2}{\partial \beta_i}
=
\langle u_2\rangle_{T_i}\langle E\rangle_{T_i}
-
\langle u_2 E\rangle_{T_i}
=
-\operatorname{Cov}_{T_i}\!\left(u_2(E),E\right).
\end{align}
Because $u_2(E)$ is a strictly increasing function of $E$, its covariance with $E$ is strictly positive:
\begin{align}
\operatorname{Cov}_{T_i}\!\left(u_2(E),E\right)>0.
\end{align}
Therefore,
\begin{align}
\frac{\partial \tilde a_2}{\partial \beta_i}<0,
\qquad
\text{or equivalently}
\qquad
\frac{\partial \tilde a_2}{\partial T_i}>0 .
\end{align}
Hence $\tilde a_2(T_i,T_b)$ is strictly monotonic in the initial temperature. Consequently, the magnitude of the slow-mode overlap cannot exhibit the nonmonotonic dependence on $T_i$ required for a weak Mpemba effect. We therefore conclude that, in the weak-dissipation limit, no final-relaxation Mpemba effect can occur in a smooth confining single-well potential for canonical initial states.

\subsection{Mpemba effects in double-well potentials} \label{sec:UD_WEak_DoubleWell}

\begin{figure}
    \centering
    \includegraphics[width=0.5\linewidth]{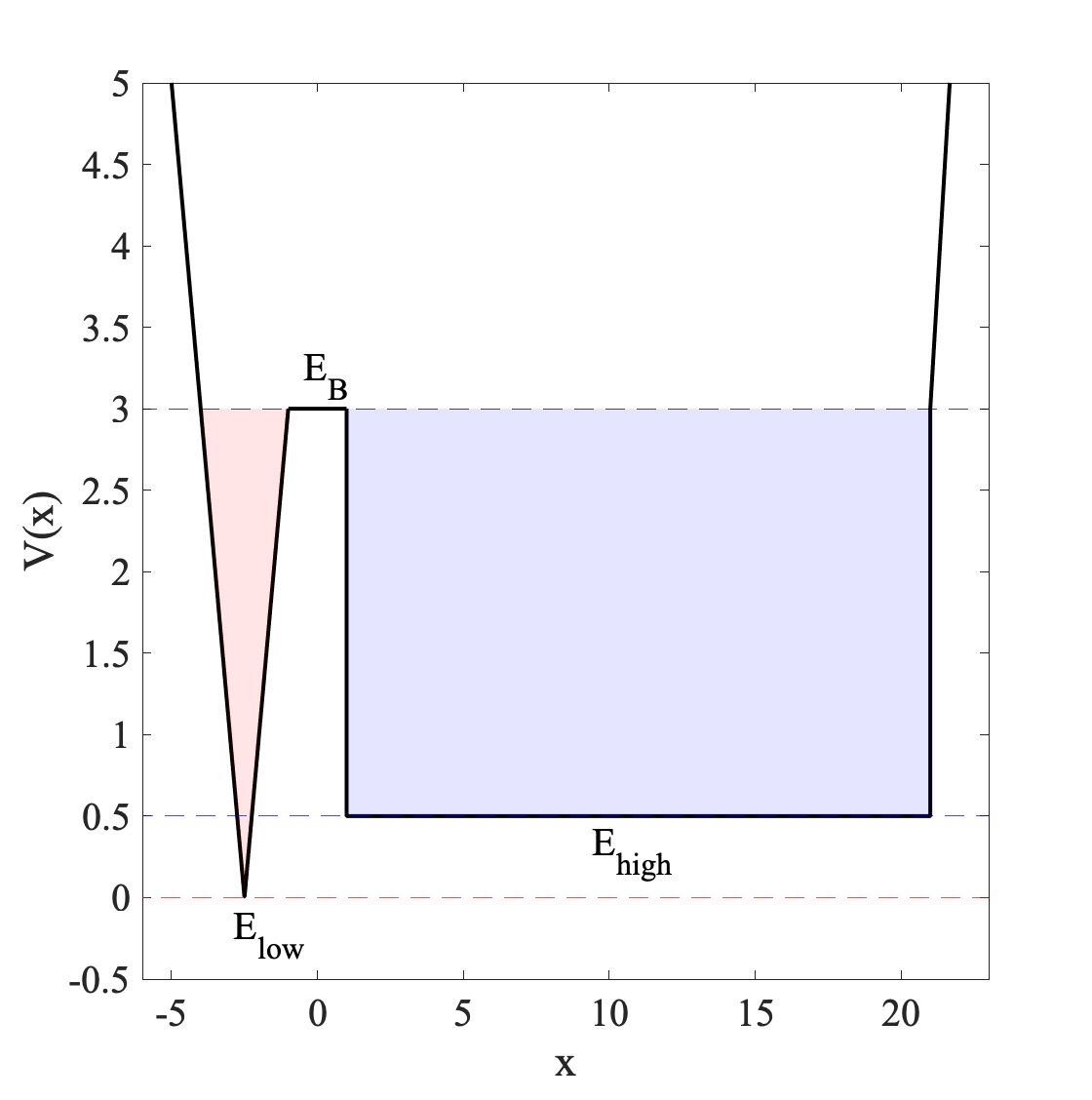}
    \caption{A double-well potential that has a strong inverse Mpemba effect even in the weak-coupling limit.}
    \label{fig:WeakDoubleWellExamp}
\end{figure}

Although in a double-well potential, the energy $E$ still serves as a 1D state variable, the situation is slightly different from the single-well case studied before. For energies below the barrier, the system cannot diffuse from an energy state in the left well to a nearby energy state in the right well without going over the barrier -- the energy diffusion in this case is limited to the specific well. Let us denote the global minimum energy by $E_\mathrm{low}$, the local minimum by $E_\mathrm{high}$, and the height of the energy barrier between them by $E_B$ (see Fig.~\ref{fig:WeakDoubleWellExamp}). Without loss of generality, we assume that the left well is deeper than the right well. 
For \(E_{\rm low}\le E<E_{\rm high}\), only the deeper well is accessible. For \(E_{\rm high}\le E<E_B\), the same energy corresponds to two disconnected orbits, one in each well. For \(E>E_B\), the orbit crosses the barrier and connects both wells.

Therefore, the mathematical description of energy diffusion in a double-well is a diffusion process on a branched state space, often referred to as a Y-graph. The system consists of three distinct branches:
(i) The left branch, $E \in [E_\mathrm{low}, E_B]$, with action $J_L(E)$ and density of states $\rho_L(E)$; (ii) The right branch, $ E \in [E_\mathrm{high}, E_B]$, with action $J_R(E)$ and density of states $\rho_R(E)$; and (iii) The top trunk, $E \in [E_B, \infty)$, with combined action $J_T(E)$ and density of states $\rho_T(E)$. Probability flux is conserved at the vertex, at $E = E_B$, where the three branches meet. 
Consequently, below the barrier, $u_2(E)$ bifurcates, taking different values on each branch (the right/left wells).  The overlap coefficient $\tilde a_2(T_i,T_b)$ is the integral of the initial canonical Boltzmann distribution against this bifurcated eigenfunction over all three branches of the energy space:
\begin{align}
\tilde a_2(T_i,T_b) \propto& \int_{E_
\mathrm{low}}^{E_B} \dd E\, u_{2,L}(E) \rho_L(E) e^{-E/k_BT_i}  + \int_{E_\mathrm{high}}^{E_B}\dd E\, u_{2,R}(E) \rho_R(E) e^{-E/k_BT_i}  \nonumber\\
& + \int_{E_B}^\infty \dd E\,u_{2,T}(E) \rho_T(E) e^{ -E/k_BT_i}\,.
\end{align}


Although a Sturm--Liouville formulation can also be introduced on the Y-graph, the single-well argument excluding the Mpemba effect no longer applies. The reason is that the energy coordinate is no longer totally ordered: below the barrier, the same value of \(E\) can correspond to distinct branches associated with the left and right wells. Consequently, \(u_2(E)\) is not a single monotone function of energy, and the different branch contributions in Eq.~(50) can have different signs. Therefore, the monotonicity argument used in the single-well case no longer guarantees that \(\tilde a_2(T_i,T_b)\) can change sign only once.

To demonstrate the \ac{ME} in this limit, we next construct a concrete example. The potential we choose is piecewise-linear, as shown in Fig.~\ref{fig:WeakDoubleWellExamp}. Such potentials were already used to study the ME in the overdamped limit~\cite{biswas2023mpemba,Walker_2021}. The left well is a `V' shape potential with a minimum at $E_\text{low}=0$, and a slope of $\pm2$. The barrier energy is $E_B=3$, and its width is 2. The right well is rectangular, with a fixed energy $E_\text{high}=0.5$ and a width of 20. Outside the wells, the potential is linear with a slope of $\pm 2$ that starts at the barrier height.   

We consider the inverse effect, with a bath temperature $T_b=1$. For any initial Boltzmann distribution with $k_BT_i<k_BT_b<E_B$, most of the distribution is concentrated either in the left well or the right well, and the slowest dynamics is the equilibration of the probability between the two wells. For a Boltzmann distribution at temperature $T_i$, the probability of being in the left well is given by
\begin{align}
    P_\text{left}(T_i) &=Z(T_i)^{-1}\int_{E_\text{low}}^{E_{B}}\dd E\,\rho_L(E)e^{-E/k_B T_i}\nonumber \\ 
&=Z(T_i)^{-1}\int_{E_\text{low}}^{E_{B}}\dd E\, 2\sqrt{2(E-E_\text{low}}) e^{-E/k_B T_i}\,,
\end{align}
where we used the density of state for the potential $V(x) = 2|x-x_\mathrm{min}|$, given by $\rho_L(E) = 2\sqrt{2m(E-E_
\mathrm{low})}$.
For the right well, the potential is constant, therefore the accumulated probability in the right well is given by:
\begin{align}
    P_\text{right}(T_i) &=Z(T_i)^{-1}\int_{E_\text{high}}^{E_{B}}\dd E\,\rho_R(E)e^{-E/k_B T_i}\nonumber \\
    &=Z(T_i)^{-1}\int_{E_\text{high}}^{E_{B}}\dd E\,\frac{L\sqrt{2}}{\sqrt{E-E_\text{high}}} e^{-E/k_B T_i}\,,
\end{align}
where we used the density of states for a flat potential with width $L$, given by $\rho_R(E) = L\sqrt{2m/(E-E_\text{high})}$.

\begin{figure}
    \centering
    \includegraphics[width=0.5\linewidth]{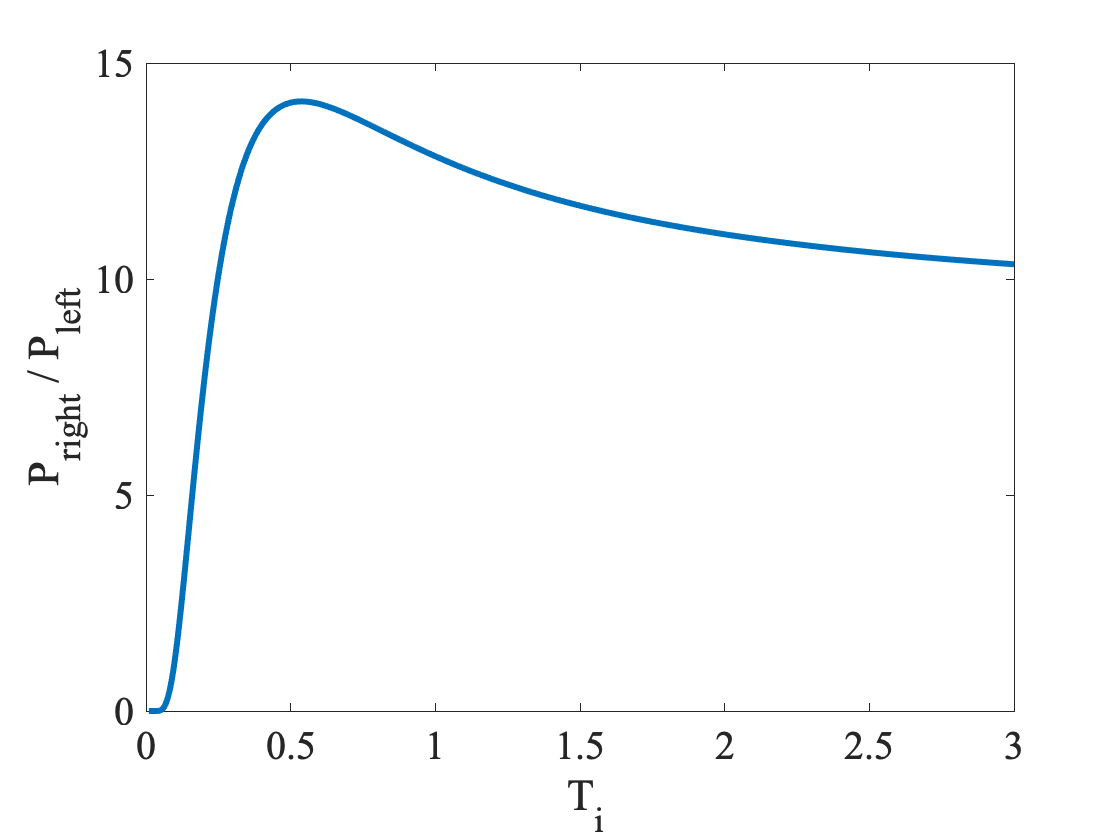}
    \caption{The ratio between the probability of being in the left well and the probability of being in the right well, for the Boltzmann distribution as a function of initial temperature $T_i$.}
    \label{fig:ProbRatio}
\end{figure}

In Fig.~\ref{fig:ProbRatio} we plot the ratio between the two, $P_\text{right}(T_i)/P_\text{left}(T_i)$ as a function of $T_i$. This ratio is nonmonotonic with $T_i$: at very low temperatures, the probability is concentrated at the lowest energy, namely in the left, `V' shaped well. As the temperature increases, the width of the right well plays a more significant role, and the large density of states near $E_\text{high}$ shifts the probability primarily to the right well. However, at even higher temperatures, higher energies are occupied. As the density of states of the `V' shape potential increases as $\sqrt{E-E_\mathrm{low}}$, whereas the density of states of the right well decreases as $1/\sqrt{E-E_\mathrm{high}}$, the probability of being in the left well grows, and the ratio decreases, generating a nonmonotonic behavior for the ratio. 

\begin{figure}
    \centering
    \includegraphics[width=0.5\linewidth]{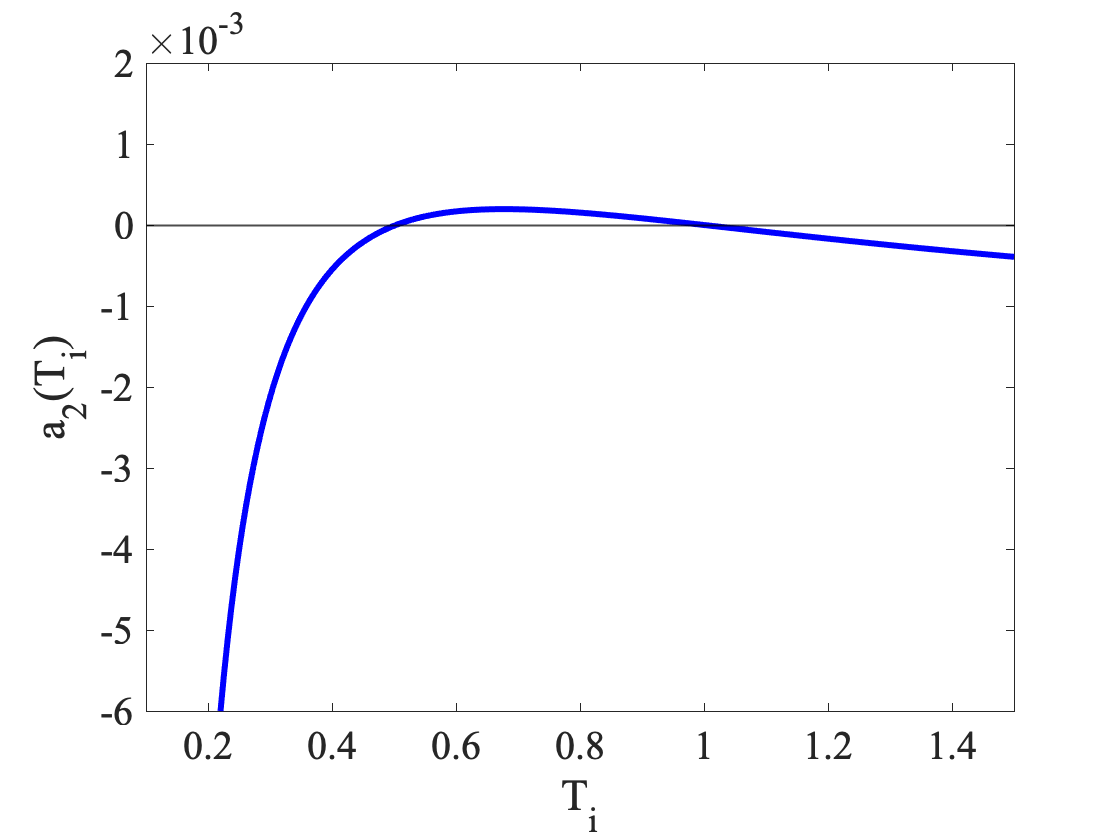}
    \caption{The coefficient of slow relaxation $\tilde a_2(T_i,T_b=1)$ for the double-well potential shown in Fig.
    ~\ref{fig:WeakDoubleWellExamp}, at $T_b=1$. As can be seen, there is a strong Mpemba effect with a zero crossing at about $T_i=0.5$.}
    \label{fig:a2_for_examp}
\end{figure}

In Fig.~\ref{fig:a2_for_examp} we plot $\tilde
a_2(T_i,T_b=1)$ for the above potential
. As can be seen, it has a strong inverse Mpemba effect. Note that the physical mechanism for this effect is quite similar to the suggested mechanism for the double well potential in the overdamped regime \cite{Lu2017Significance,Kumar2022Anomalous}. However, in the overdamped regime, entropy is dominated by the well width, whereas in the weak-damping regime it is dominated by the density of states. This difference is significant: the potential considered in Fig.~\ref{fig:potential}, which exhibits the \ac{ME} in the overdamped case, does not exhibit any type of \ac{ME} in the weak-damping limit. However, the potential in Fig.~\ref{fig:WeakDoubleWellExamp} exhibits a strong, inverse \ac{ME} in the weak-damping limit due to different scalings of density of states between the left and right wells. This potential actually has a strong inverse Mpemba effect also in the overdamped limit. 

Finally, note that the potential in Fig.~\ref{fig:WeakDoubleWellExamp} should be understood as an idealized, analytically tractable limiting potential, introduced to highlight the different scalings of the density of states in the left and right wells. The same mechanism is expected to persist under sufficiently small smooth regularizations of the corners and vertical walls. More broadly, this limiting example suggests that ~\ac{ME} may also arise in other weak-damping scenarios in which the two wells of the double-well potential exhibit sufficiently distinct energy-density scalings.

\section{Conclusions} \label{sec:conclusions}

Our results show that final-relaxation Mpemba effects can survive inertial dynamics. Near the overdamped limit, this persistence follows perturbatively. In the strict weak-damping limit, the effect is absent for single-well potentials with canonical initial states, but can arise in multiwell systems whose energy-space dynamics is branched.

We numerically demonstrate examples in which Mpemba effects survive the inclusion of inertia, extending all the way to the weak-damping limit. In particular, our results show that varying the dissipation rate while keeping the temperature fixed can modify both the relaxation timescale and the qualitative character of the relaxation process. For example, the system may transition from a weak direct effect to a strong direct effect. In this case, increasing the damping not only accelerates relaxation, but also changes the direction from which the system approaches equilibrium.

It is instructive to compare the weak-damping and overdamped limits. In the overdamped limit, the \ac{ME} can occur even in a single-well potential, whereas in the weak-damping limit a barrier is essential for the effect to arise. In a double-well potential, however, the two limits share the same slowest relaxation mechanism: hopping over the barrier, or equivalently, balancing the probability between the two wells. This mechanism can give rise to the Mpemba effect in both regimes, although the entropic contributions associated with each well differ between the overdamped and weak-damping limits.

Exploring the underdamped regime brings us one step closer to understanding the Mpemba effect in water, although the path toward this goal is likely to remain challenging. Even if a final-relaxation Mpemba effect exists in an effective \ac{1D} double-well potential associated with hydrogen and covalent bonds, the emergence of an analogous effect in the full many-body system is not guaranteed. This is even more true for the Mpemba effect through a first-order phase transition, as observed in water. Nevertheless, we hope that connections between these simplified descriptions and water's behavior will emerge in future research.

\begin{acknowledgments}
    M.~V. acknowledges the hospitality of MPI-PKS. This material is based on work supported by the National Science Foundation under Grant No.~DMR-1944539. S.~A.~S. is supported by the CHE/PBC Fellowship. O.~R. acknowledges financial support from ISF Grant No. 232/23,  the Minerva foundation with funding from the Federal German Ministry for Education and Research and the Ministry of Research and Culture in Lower Saxony and the Volkswagen Foundation.
\end{acknowledgments}

\bibliography{references}

\end{document}


\title{
Supplemental Material: Extending the Mpemba effect to the underdamped realm
}
\author{Shahaf Aharony Shapira}
\email{shahaf.aharony@weizmann.ac.il}
\affiliation{Department of Physics of Complex Systems, Weizmann Institute of Science, Rehovot 7610001, Israel}

\author{Gene Chen}
\email{wkp5pp@virginia.edu}
\affiliation{Department of Physics, University of Virginia, Charlottesville, VA 22904, USA}

\author{Marija Vucelja}
\email{mvucelja@virginia.edu}
\affiliation{Department of Physics, University of Virginia, Charlottesville, VA 22904, USA}
\affiliation{Department of Mathematics, University of Virginia, Charlottesville, VA 22904, USA}
\affiliation{Max Planck Institute for the Physics of Complex Systems, N\"othnitzer Str. 38, Dresden, 01187, Germany}

\author{Oren Raz}
\email{oren.raz@weizmann.ac.il}
\affiliation{Department of Physics of Complex Systems, Weizmann Institute of Science, Rehovot 7610001, Israel}

\maketitle
\onecolumngrid
\tableofcontents

\begin{acronym}
    \acro{ME}{Mpemba effect}
    \acro{DoF}{degree of freedom}
    \acro{NTE}{negative thermal expansion}
    \acro{KK}{Klein-Kramers}
    \acro{SM}{Supplemental Material}
    \acro{PDF}{probability density function}
    \acro{MCF}{Matrix Continued Fractions}
    \acro{1D}{one-dimensional}
    \acro{2D}{two-dimensional}
    \acro{PDE}{partial derivative equation}
\end{acronym}

\section{Strong dissipation limit} 

The \ac{KK} equation can be written as 
\begin{align}
    \partial _t f(x,p,t) = \mathcal{L}(x,p)f(x,p,t), 
\end{align}
here $f(x,p,t)$ is the probability density function of finding a particle at coordinate $x$ and with momentum $p$ at time $t$, and $\mathcal{L}(x,p)$ is the forward \ac{KK} generator
\begin{align}
    \mathcal{L}(x,p) = - \frac{p}{m} \partial_x + \partial_p \left(V' + \frac{\gamma}{m}p\right) + \gamma k_B T_b \partial ^2 _p, 
\end{align}
where $m$ is the mass of the particle, $V(x)$ is the external potential, with $V'(x) \equiv \dd V(x) / \dd x$, $\gamma$ is the damping constant, $k_B$ is the Boltzmann constant and $T_b$ is the bath temperature. The Hamiltonian $\mathcal{H}(x,p)$ is 
\begin{align}
\label{eq:Hamiltonian}
    \mathcal{H}(x,p) = \frac{p^2}{2 m} + V(x). 
\end{align}
The equilibrium distribution is the Boltzmann distribution at temperature $T_b$, 
\begin{align}
\label{eq:Boltzmann-dist}
    f_{\rm eq} (x,p,T_b) = \frac{1}{\mathcal{Z}(T_b)} e^{ - \frac{\mathcal{H}(x,p)}{k_BT_b}}\,, 
\end{align}
where $\mathcal{Z}$ is the partition function
\begin{align}
    \mathcal{Z}(T_b) = \int _{\mathcal{D}_x} \dd x \int ^\infty _{-\infty} \dd p \,e^{ - \frac{\mathcal{H}(x,p)}{k_BT_b}}\,, 
\end{align}
where $\mathcal{D}_x$ is the spatial domain. The spatial domain here can be finite or infinite.
The forward \ac{KK} generator can be  written as 
\begin{align}
    \mathcal{L} f &= \{\mathcal{H}, f \} + \epsilon \mathcal{H}_1f\,, 
\end{align}
with the $\{\,,\}$ denote the Poisson brackets. Here $\epsilon \equiv \gamma/(m \omega_0)$ and $\omega_0$ is the natural frequency of the unperturbed Hamiltonian. We elaborate how $\omega_0$ can be chosen in the main text. The derivations below do not depend on the specific choice of $
\omega_0$. The term $\mathcal{H}_1$ is the ``dissipative'' part, that specifies the connection to the Markovian bath. 
Note that $\mathcal{H}_1$ is the Ornstein-\"Uhlenbeck operator in momentum 
\begin{align}
    \mathcal{H}_1 f &= \omega_0\partial_p \left(p f + D_p(T_b) \partial _p f\right), 
\end{align}
with equilibrium momentum variance $D_p(T_b) \equiv mk_B T_b$. 
The nullspace of $\mathcal{H}_1$, $\mathcal{H}_1 M(p,T_b) = 0$, is a Gaussian in momentum 
\begin{align}
\label{eq:Gaussian-momentum}
M(p,T_b) = \frac{1}{\sqrt{2\pi D_p(T_b)}}e^{-\frac{p^2}{2 D_p(T_b)}}.  
\end{align}
The scalar product is defined as 
\begin{align}
    \langle g,f \rangle \equiv \int _{\mathcal{D}_x}\dd x \, \int^\infty _{-\infty}   \dd p \, g(x,p) f (x,p)\,.
\end{align}
The system obeys reflective (natural) boundary conditions, with the probability current density going to zero at (the boundaries of the spatial domain) infinity. 

The corresponding adjoint of $\mathcal{H}_1$ is 
\begin{align}
\label{eq:H1dagger}
\mathcal{H}_1^\dagger = \omega _0 \left( - p \partial _p + D_p(T_b) \partial _p ^2 \right). 
\end{align}
A constant function is in its nullspace
\begin{align}
    \mathcal{H}_1 ^\dagger 1 = 0. 
\end{align}

The momentum relaxation timescale is $\tau_p \propto m/\gamma$, while the coordinate relaxation timescale is $\tau _x \propto \omega_0 ^{-1}$. 
A relevant parameter is 
the ratio of two timescales 
\begin{align}
\label{eq:epsilon}
    \epsilon \equiv \frac{\tau_x}{\tau_p} = \frac{\gamma}{m\omega_0}. 
\end{align}
In the limit of high friction
\begin{align}
    \epsilon \gg 1. 
\end{align}
The limit is characterized by strong dissipation, fast momentum thermalization compared to a slow diffusion of the coordinate. In contrast, the low-friction regime is characterized by slow energy relaxation, allowing oscillations of the system. 
In high-friction regimes, the momentum is often integrated over in order to focus on the slow dynamics in coordinate space. In the following we look at the eigenvalue problem of the \ac{KK} operator and its adjoint in orders of $1/\epsilon$, for more details, see, e.g.,~\cite{gardiner2009stochastic}.

\subsection{Right eigenfunctions}

Let us look at the following eigenvalue problem 
\begin{align}
    \mathcal{L}\,v_n(x,p) = \lambda_n v_n(x,p), 
\end{align}
in the strong dissipation limit, $\epsilon \gg 1$.  The relaxation in momentum is fast, the slow eigenmodes, however, have a slow branch tied to the Smoluchowski (overdamped Langevin) dynamics and fast branches separated by $\mathcal{O}(\epsilon)$ for large damping. The slow underdamped right eigenfunction is thus approximately 
\begin{align}
    v_n(x,p) = v ^{(0)}_n(x,p) + \frac{1}{\epsilon} v^{(1)} _n (x,p) + \frac{1}{\epsilon ^2}v_n ^{(2)}(x,p) + \mathcal{O}\left(\epsilon^{-3}\right), 
\end{align}
and the ``slow'' eigenvalues are
\begin{align}
    \lambda_n = \frac{\mu_n}{\epsilon}  + \mathcal{O}\left(\epsilon^{-3}\right). 
\end{align}
Note that the next correction to the eigenvalue is proportional to $\epsilon^{-3}$, as due to the parity of the momentum correction in $v_n(x,p)$ the terms of order $\epsilon^{-2}$ are zero. 

Therefore, the \ac{KK} equation is 
\begin{align}
\nonumber
    \left(\frac{\mu _n}{\epsilon} + \mathcal{O}\left(\epsilon^{-3}\right) \right) \left(v^{(0)} _n + \frac{1}{\epsilon}v^{(1)} _n + \mathcal{O}\left(\epsilon^{-2}\right) \right)=& \left\{\mathcal{H},\left(v^{(0)} _n + \frac{1}{\epsilon}v^{(1)} _n +  \mathcal{O}\left(\epsilon^{-2}\right) \right)\right\} 
    \\&+ \epsilon \mathcal{H}_1 \left(v^{(0)} _n + \frac{1}{\epsilon}v^{(1)} _n + \frac{1}{\epsilon^2}v^{(2)} _n + \mathcal{O}\left(\epsilon^{-3}\right) \right)\,. 
\end{align}
This gives orders of $\epsilon$, $1$, and $\epsilon^{-1}$:
\begin{subequations}
\begin{align}
0 &= \mathcal{H}_1 v_n ^{(0)},
\\
0 &= \left\{\mathcal{H}, v_n ^{(0)}\right\} + \mathcal{H}_1 v_n ^{(1)},
\\
\mu _n v_n ^{(0)} &=  \left\{\mathcal{H}, v_n ^{(1)}\right\} +  \mathcal{H}_1 v^{(2)} _n. 
\end{align}
\end{subequations}
The leading order is obtained from the nullspace of $\mathcal{H}_1$,  i.e., from $\mathcal{H}_1v_n(x,p) = 0$, and it is 
\begin{align}
\label{eq:f0}
    v_n ^{(0)}(x,p) = M(p,T_b) v_n^{(S)} (x), 
\end{align}
where $M(p,T_b)$ is given by Eq.~\eqref{eq:Gaussian-momentum} and $v_n^{(S)}(x)$ is the overdamped eigenfunction.  
The first order correction is 
\begin{align}
    \left\{\mathcal{H},v_n ^{(0)}\right\} +   \mathcal{H}_1 v_n ^{(1)} = 0. 
\end{align}
Using, in the above equation, the Hamiltonian $\mathcal{H}(x,p)$ from Eq.~\eqref{eq:Hamiltonian} and $v_n ^{(0)}(x,p)$ from Eq.~\eqref{eq:f0} we have
\begin{align}
 - p M(p,T_b) \left(\frac{V'(x)}{D_p(T_b)}   v_n^{(S)}(x)+ \frac{1}{m}\left(v_n^{(S)}(x)\right)'\right) + \mathcal{H}_1v_n ^{(1)}(x,p)  = 0. 
\end{align}
Using the fact that $\mathcal{H}_1$ acts only on the momentum degrees of freedom, and that
\begin{align}
    \mathcal{H}_1\, p M(p,T_b) = - \omega _0 p M(p,T_b),
\end{align}
we have 
\begin{align}
 \mathcal{H}_1 \frac{1}{\omega_0} p M(p,T_b) \left(\frac{V'(x)}{D_p(T_b)}   v_n^{(S)}(x)+ \frac{1}{m}\left(v_n^{(S)}(x)\right)'\right) + \mathcal{H}_1v_n ^{(1)}(x,p)  = 0. 
\end{align}
Therefore one solution of the above equation is 
\begin{align}
\label{eq:f1}
    v_n ^{(1)}(x,p)=- \frac{p}{\omega_0} M(p,T_b) \left(\frac{V'(x)}{D_p(T_b)}v_n^{(S)}(x) + \frac{1}{m}\left(v_n^{(S)}(x)\right)' \right) + B_n(x).  
\end{align}
Note that $B_n(x)$, since it is independent of momentum, would change the slow mode. We do not want that and hence we demand that 
\begin{align}
\int ^\infty _{-\infty} \dd p \, v_n ^{(1)}(x,p) = 0, 
\end{align}
and thus we have 
\begin{align}
    B_n(x) = 0. 
\end{align}
Hence $v_n ^{(1)}(x,p)$ according to Eq.~\eqref{eq:f1} is
    \begin{align}
\label{eq:f1_final-a}
 v_n ^{(1)}(x,p)&=- \frac{p}{\omega_0 m} M(p,T_b) A_n(x), 
\end{align}
with 
\begin{align}
 \label{eq:f1_final-b}
    A_n(x) &\equiv \left(\frac{V'(x)}{k_B T_b}v_n^{(S)}(x) + \left(v_n^{(S)}(x)\right)' \right). 
\end{align}
Therefore, the first-order correction to the eigenfunction is 
\begin{align}
    v_n(x,p) = M(p,T_b) \left(v_n^{(S)}(x) -\frac{1}{\epsilon} \frac{p}{\omega_0m} A_n(x)\right) + \mathcal{O}\left(\epsilon^{-2}\right).
\end{align}
The next order of \ac{KK} equation, with the $\mathcal{O}(\epsilon^{-1})$ terms, obeys the following expression
\begin{align}
    \left\{\mathcal{H}, v_n ^{(1)}\right\} + \mathcal{H}_1v^{(2)} _n = \mu _n v_n ^{(0)}. 
\end{align}
Plugging in Eqs.~\eqref{eq:Hamiltonian},~\eqref{eq:f0} and~\eqref{eq:f1_final-a} we get 
\begin{align}
\label{eq:f2_eq}
    V'(x)\left(-\frac{1}{m\omega_0} A_n(x)\right)\left(M(p,T_b) -\frac{p^2}{D_p(T_b)}M(p,T_b)\right) + \frac{p^2}{m^2\omega _0} M(p,T_b) A_n'(x) + \mathcal{H}_1 v^{(2)} _n = \mu _n M(p,T_b)v_n^{(S)}(x). 
\end{align}
We impose solvability by demanding that the next order correction is orthogonal to the nullspace of $\mathcal{H}_1$, i.e., 
\begin{align}
    \int ^\infty _{-\infty}\dd p \,\mathcal{H}_1 v^{(2)} _n(x,p) =0. 
\end{align}

Therefore, integrating over momentum, we get 
\begin{align}
\nonumber
    &V'(x)\left(-\frac{1}{m\omega_0} A_n(x)\right) \int ^\infty _{-\infty}\dd p \,\left(M(p,T_b) -\frac{p^2}{D_p(T_b)}M(p,T_b)\right) +\int ^\infty _{-\infty}\dd p \, \frac{p^2}{m^2\omega _0} M(p,T_b) A_n'(x) 
    \\
    &= \mu _n \int ^\infty _{-\infty}\dd p \, M(p,T_b) v_n^{(S)}(x),
\end{align}
which reduces to 
\begin{align}
\label{eq:Lsmol-aux}
    &\frac{D_p(T_b)}{m^2\omega _0} A_n'(x) 
    = \mu _n v_n^{(S)}(x). 
\end{align}
Plugging in the expression for $A_n(x)$, from Eq.~\eqref{eq:f1_final-b}, we get
\begin{align}
    &\frac{1}{m\omega _0} \partial _x \left(V'(x) v_n^{(S)} (x) +k_B T_b\left( v_n^{(S)}(x) \right)'\right)  
    = \mu _n v_n^{(S)}(x). 
\end{align}
Using $\epsilon$, from Eq.~\eqref{eq:epsilon}, we have 
\begin{align}
\label{eq:Smoluchowski}
\mathcal{L}^{(S)}v_n^{(S)}(x)  
= \frac{\mu _n}{\epsilon} v_n^{(S)}(x), 
\end{align}
where $\mathcal{L}_S$ is the forward Smoluchowski (overdamped Fokker-Planck) generator
\begin{align}
\mathcal{L}_S \equiv \frac{1}{\gamma}\partial _x \left(V'(x) + k_B T_b\partial_x\right).
\end{align}
This is consistent with $v_n^{(S)}(x)$ as the right eigenfunction of $\mathcal{L}_S$ with the corresponding eigenvalue 
\begin{align}
\label{eq:eig-Smo-eps}
     \lambda^{(S)}_n = \frac{\mu _n}{\epsilon}.
\end{align}

The next order, in $
\epsilon$, of \ac{KK} equation, is obtained by solving  Eq.~\eqref{eq:f2_eq} for $v^{(2)} _n$, 
\begin{align}
\mathcal{H}_1 v^{(2)} _n = \mu _n M(p,T_b)v_n^{(S)}(x) - V'(x)\left(-\frac{1}{m\omega_0} A_n(x)\right)\left(1 -\frac{p^2}{D_p(T_b)}\right)M(p,T_b) - \frac{p^2}{m^2\omega _0} M(p,T_b) A_n'(x)
\end{align}
By plugging in Eq.~\eqref{eq:Lsmol-aux}, we get 
\begin{align}
\label{eq:f2_eq-01}
\mathcal{H}_1 v^{(2)} _n =  -\frac{1}{m^2\omega_0} \left(p^2-D_p(T_b)\right)M(p,T_b)\left(\frac{V'(x)}{k_BT_b}A_n(x) + A_n'(x)\right).
\end{align}
Using 
\begin{align}
    \mathcal{H}_1 (p^2 - D _p(T_b)) M(p,T_b) = - 2 \omega _0 (p^2 - D_p(T_b)) M(p,T_b),
\end{align}
in Eq.~\eqref{eq:f2_eq-01} we get 
\begin{align}
\mathcal{H}_1 v^{(2)} _n =  \frac{1}{2m^2\omega_0^2} \mathcal{H}_1\left(p^2-D_p(T_b)\right)M(p,T_b)\left(\frac{V'(x)}{k_BT_b}A_n(x) + A_n'(x)\right). 
\end{align}
Hence, a solution for $v^{(2)} _n$ is 
\begin{align}
\label{eq:f2_final}
    v^{(2)} _n(x,p) = \frac{1}{2m^2\omega_0^2} \left(p^2-D_p(T_b)\right)M(p,T_b)\left(\frac{V'(x)}{k_BT_b}A_n(x) + A_n'(x)\right) + G_n(x). 
\end{align}
Without loss of generality, we will set the gauge freedom 
\begin{align}
    G_n(x) = 0, 
\end{align}
and choose a gauge when defining the corresponding correction to the left eigenfunction so that the scalar product of the left and right eigenfunctions corresponding to the same eigenvalue $\lambda_n$, is biorthogonal, 
\begin{align}
\langle u_n, v_m \rangle = \delta_{nm},      
\end{align}
for $\mathcal{L} \, v_n = \lambda_n v_n$ and $\mathcal{L}^\dagger\, u_n = \lambda_n u_n$. Here, $\mathcal{L}^\dagger$ is the adjoint operator of $\mathcal{L}$. 

Finally, the right eigenfunction is 
\begin{align}
\label{eq:f-O3}
    v_n(x,p) = M(p,T_b) \left(v_n^{(S)}(x) -\frac{1}{\epsilon} \frac{p}{\omega_0m} A_n(x) +  \frac{1}{\epsilon^2}\frac{1}{2m^2 \omega _0^2}(p^2 - D_p(T_b))C_n(x)\right) + \mathcal{O} \left(\epsilon^{-3}\right), 
\end{align}
where according to Eqs.~\eqref{eq:f1_final-b} and~\eqref{eq:f2_final} the functions $A_n(x)$ and $C_n(x)$ are 
\begin{align}
 A_n(x) &\equiv \left(\frac{V'(x)}{k_B T_b}v_n^{(S)}(x) + \left(v_n^{(S)}(x)\right)'
 \right), 
\\
    C_n(x) &\equiv \left(\frac{V'(x)}{k_BT_b}A_n(x) + A_n'(x)\right). 
\end{align}

\subsection{Left eigenfunctions}

Analogously one can approximate the left eigenfunctions, which are solutions to the following 
equation 
\begin{align}
\mathcal{L}^\dagger \, u_n(x,p) = \lambda_n u_n(x,p). 
\end{align}
The adjoint KK operator is 
\begin{align}
    \mathcal{L}^\dagger u_n = -\{\mathcal{H},u_n\} + \epsilon \mathcal{H}^\dagger _1 u_n, 
\end{align}
where $\mathcal{H}^\dagger$ is defined in Eq.~\eqref{eq:H1dagger}. 

The ansatz for the eigenfunction $u_n(x,p)$ in the strong dissipation limit is 
\begin{align}
    u_n(x,p) = u^{(0)} _n (x,p) + \frac{1}{\epsilon} u^{(1)} _n (x,p) + \frac{1}{\epsilon ^2}u^{(2)}_n(x,p) + \mathcal{O}\left(\epsilon^{-3}\right),
\end{align}
The equation with $\epsilon-$order terms is 
\begin{align}
    \mathcal{H}_1 ^\dagger u^{(0)} _n(x,p)  =0\,. 
\end{align}
Therefore, the momentum part of the left eigenfunction is a constant, and the left eigenfunction is  
\begin{align}
\label{eq:g0}
    u^{(0)} _n (x,p) = u_n^{(S)}(x)\,, 
\end{align}
where $u_n^{(S)}(x)$ is the left eigenfunction of $\mathcal{L}_S$. The $\epsilon^{0}-$order gives the following equation
\begin{align}
    -\left\{\mathcal{H},u^{(0)} _n \right\} + \mathcal{H}_1^\dagger u^{(1)} _n  = 0. 
\end{align}
Substituting $u^{(0)} _n(x,p) $, from Eq.~\eqref{eq:g0}, we get 
\begin{align}
\frac{p}{m}\left(u_n^{(S)}(x)\right)' + \mathcal{H}_1^\dagger u^{(1)} _n  = 0.
\end{align}
Using 
\begin{align}
    \mathcal{H}^\dagger _1 p = - \omega _0 p, 
\end{align}
we get 
\begin{align}
 \mathcal{H}_1^\dagger u^{(1)} _n(x)  = \mathcal{H}^\dagger _1 \frac{p}{m\omega _0}\left(u_n^{(S)}(x)\right)'. 
\end{align}
Hence the solution is 
\begin{align}
\label{eq:g1}
    u^{(1)} _n (x,p) = \frac{p}{m\omega_0}\left(u_n^{(S)}(x)\right)'+ \tilde B_n(x), 
\end{align}
where $\tilde B_n(x)$ should be set such that it is orthogonal to the slow manifold
\begin{align}
    \int ^\infty _{-\infty} \dd p \, \tilde B_n(x) M(p,T_b) = 0, 
\end{align}
which leads to 
\begin{align}
    \tilde B_n(x) = 0. 
\end{align}
The term $\tilde B_n(x)$ lies in the nullspace of $\mathcal{H}_1^\dagger$ and corresponds to a gauge freedom in the definition of the slow component. We fix this freedom by requiring the first-order correction to have zero Maxwellian average,
\[
\int_{-\infty}^{\infty} dp\,M(p,T_b)\,u_n^{(1)}(x,p)=0,
\]
which gives $\tilde B_n(x)=0$.
Thus we have 
\begin{align}
\label{eq:g1-final}
    u^{(1)} _n (x,p) = \frac{p}{m\omega_0}\left(u_n^{(S)}(x)\right)'\,. 
\end{align}

The expression for the approximate left eigenfunction up to $\mathcal{O}(\epsilon^{-2})$ terms is 
\begin{align}
    u_n(x,p) = u_n^{(S)}(x) + \frac{1}{\epsilon}\frac{p}{m\omega _0}\left(u_n^{(S)}(x)\right)' + \mathcal{O}\left(\epsilon^{-2}\right).
\end{align}
The $\epsilon^{-1}-$order is  
\begin{align}
- \left\{\mathcal{H},u^{(1)} _n \right\} + \mathcal{H}_1^\dagger u^{(2)} _n  = \mu _n u^{(0)} _n . 
\end{align}
Substituting the expressions for $u^{(0)} _n $ and $u^{(1)} _n $ from Eqs.~\eqref{eq:g0} and~\eqref{eq:g1-final} we have 
\begin{align}
\label{eq:g2-exp}
    -V'(x)\frac{1}{m\omega_0}\left(u_n^{(S)}(x)\right)' + \frac{p^2}{m^2 \omega _0}\left(u_n^{(S)}(x)\right)'' + \mathcal{H}_1^\dagger u^{(2)} _n  = \mu _n u_n^{(S)}(x). 
\end{align}
For solvability the $u^{(2)} _n $ term should be orthogonal to the null space of $\mathcal{H}_1$, hence $u^{(2)} _n $ should obey 
\begin{align}
    \int ^\infty _{-\infty}\dd p \, M(p,T_b) \mathcal{H}^\dagger _1 u^{(2)} _n  (x,p) = 0. 
\end{align}
After multiplying Eq.~\eqref{eq:g2-exp} with $M(p,T_b)$ and integrating over momentum, we obtain the following
\begin{align}
    -V'(x)\frac{1}{m\omega_0}\left(u_n^{(S)}(x)\right)' + \frac{D_p(T_b)}{m^2 \omega _0}\left(u_n^{(S)}(x)\right)'' = \mu _n u_n^{(S)}(x)\,.
\end{align}
The above expression can be written as  
\begin{align}
    \mathcal{L}^\dagger _S u_n^{(S)}(x) = \frac{1}{\gamma}\left(-V'(x)\partial_x + k_BT_b\partial ^2 _x \right)u_n^{(S)}(x) = \frac{\mu _n}{\epsilon}u_n^{(S)}(x)\,,  
\end{align}
where $\mathcal{L}_S^\dagger$ is the adjoint Smoluchowski generator, with eigenvalue problem 
\begin{align}
     \mathcal{L}^\dagger _S \,u_n^{(S)}(x) = \lambda^{(S)} _n  u_n^{(S)}(x). 
\end{align}
Hence $u_n^{(S)}(x)$ is the left eigenfunction of the Smoluchowski generator with eigenvalue $\lambda^{(S)}_n = \mu _n/\epsilon$, which is consistent with our ansatz, Eq.~\eqref{eq:g0}. 

Plugging in the overdamped equation for the backward Smoluchowski generator, $\mathcal{L}^\dagger _{S}$, in Eq.~\eqref{eq:g2-exp}, we get 
\begin{align}
     \mathcal{H}_1^\dagger u^{(2)} _n  = -\frac{1}{m^2 \omega _0}\left(p^2 -D_p(T_b)\right)\left(u_n^{(S)}(x)\right)''.
\end{align}
Using 
\begin{align}
    \mathcal{H}_1^\dagger\left(p^2 -D_p(T_b)\right) = - 2 \omega _0 \left(p^2 -D_p(T_b)\right),
\end{align}
we get 
\begin{align}
      \mathcal{H}_1^\dagger u^{(2)} _n  = \frac{1}{2 m^2 \omega _0^2 } \mathcal{H}_1^\dagger\left(p^2 -D_p(T_b)\right)\left(u_n^{(S)}(x)\right)''. 
\end{align}
The solution of the above equation is 
\begin{align}
\label{eq:g2-a}
    u^{(2)} _n (x,p) = \frac{1}{2 m^2 \omega _0^2 } \left(p^2 -D_p(T_b)\right)\left(u_n^{(S)}(x)\right)'' + F_n(x), 
\end{align}
where $F_n(x)$ can be determined by demanding that the biorthogonality of the left and the right eigenfunction holds, that is
\begin{align}
    \int _{\mathcal{D}_x} \dd x \,\int ^\infty _{-\infty} \dd p\, u_n(x,p) v_n(x,p) = 1 + \mathcal{O}\left(\epsilon^{-3}\right). 
\end{align}
Finally, the left eigenfunction $u_n(x,p)$ using Eqs.~\eqref{eq:g0},~\eqref{eq:g1-final} and~\eqref{eq:g2-a} is 
\begin{align}
\label{eq:g-O3}
    u_n(x,p) = u_n^{(S)}(x) + \frac{1}{\epsilon}\frac{p}{m\omega _0}\left(u_n^{(S)}(x)\right)' + \frac{1}{\epsilon^2}\left(\frac{1}{2 m^2 \omega _0^2 } \left(p^2 -D_p(T_b)\right)\left(u_n^{(S)}(x)\right)'' + F_n(x)\right)+\mathcal{O}\left(\epsilon^{-3}\right). 
\end{align}
Applying in the scalar product $\langle u_n, v_n\rangle$ the expressions for $u_n$ and $v_n$, Eqs.~\eqref{eq:f-O3} and~\eqref{eq:g-O3} we get 
\begin{align}
    \langle u_n, v_n\rangle = 1 + \frac{1}{\epsilon^2}\left( \int  _{\mathcal{D}_x}\dd x \, F_n(x) v_n^{(S)} (x) - \int  _{\mathcal{D}_x}\dd x \, \frac{D_p(T_b)}{m^2 \omega _0^2}\left(u_n^{(S)}(x)\right)' A_n(x)\right) + \mathcal{O}\left(\epsilon^{-3}\right).   
\end{align}

Using integration by parts and the boundary conditions (where the probability density current is zero at the boundary, i.e., $A_n(
\partial\mathcal{D}_x
)  = 0$) and Eq.~\eqref{eq:Lsmol-aux} we get
\begin{align}
     \int _{\mathcal{D}_x} \dd x \, F_n(x) v_n^{(S)} (x) =- \int  _{\mathcal{D}_x} \dd x \, \frac{\mu _n}{\omega_0} u_n^{(S)}(x) v_n^{(S)}(x), 
\end{align}
we have 
\begin{align}
     F_n(x) = -\frac{\mu _n}{\omega_0} u_n^{(S)}(x). 
\end{align}
Therefore, after plugging $F_n(x)$ in the expression for $u^{(2)}_n(x,p)$, Eq.~\eqref{eq:g2-a}, we get
\begin{align}
    u^{(2)} _n (x,p) = \frac{1}{2 m^2 \omega _0^2 } \left(p^2 -D_p(T_b)\right)\left(u_n^{(S)}(x)\right)'' - \frac{\mu _n}{\omega_0} u_n^{(S)}(x). 
\end{align}
Hence, the left eigenfunction is 
\begin{align}
\label{eq:g-O3-final}
    u_n(x,p) = u_n^{(S)}(x) + \frac{1}{\epsilon}\frac{p}{m\omega _0}\left(u_n^{(S)}(x)\right)' + \frac{1}{\epsilon^2}\left(\frac{1}{2 m^2 \omega _0^2 } \left(p^2 -D_p(T_b)\right)\left(u_n^{(S)}(x)\right)'' - \frac{\epsilon \lambda^{(S)}_n}{\omega_0} u_n^{(S)}(x)\right)+\mathcal{O}\left(\epsilon^{-3}\right). 
\end{align}

\subsection{Overlap coefficient}

The overlap coefficient of the largest nonzero mode is defined as the normalized projection of the corresponding left eigenvector to the initial condition. For the case where the initial condition is a Boltzmann distribution at temperature $T_i$, the overlap is 
\begin{align}
    a_2(T_i,T_b) = \int _{\mathcal{D}_x} \dd x \int ^\infty _{-\infty} \dd p \,  u _2 (x,p) f_{\rm eq}(x,p;T_i)\,,
\end{align}
where we use the biorthogonal normalization $\langle u_2,v_2\rangle=1$. 
Substituting the left eigenvector $u_2(x,p)$, from  Eq.~\eqref{eq:g-O3-final}, and the Boltzmann distribution at $T_i$ in the definition of the overlap $a_2$ we get 
\begin{align}
\label{eq:a2-under}
a_2(T_i,T_b) = a_2^{(S)}(T_i,T_b) + 
    \frac{1}{\epsilon^2}\left(\frac{1}{2 m^2 \omega_0 ^2}\left(D_p(T_i) - D_p(T_b)
    \right) \left\langle\left(u_2 ^{(S)}(x)\right)''\right\rangle_{f^{(S)}_\mathrm{eq}(T_i)} - \frac{\epsilon \lambda_2^{(S)}}{\omega_0}a^{(S)}_2(T_i,T_b)\right) 
+\mathcal{O}\left(\epsilon^{-3}\right)\,, 
\end{align}
where $a_2^{(S)}(T_i, T_b)$ is the corresponding overlap in the overdamped case, and 
\begin{align}
    \left\langle\left(u_2 ^{(S)}(x)\right)''\right\rangle_{f^{(S)}_\mathrm{eq}(T_i)} 
\equiv \int _{\mathcal{D}_x} \dd x \,f_\mathrm{eq} (x,T_i) \left(u_2 ^{(S)}(x)\right)''\,,  
\end{align}
is the expectation value of $\left(u_2 ^{(S)}(x)\right)''$ with respect to the Boltzmann distribution at $T_i$ in the overdamped case 
\begin{align}
    f_\mathrm{eq}^{(S)}(x,T_i) = \frac{1}{\mathcal{Z}^{(S)}(T_i)}e^{- \frac{V(x)}{k_BT_i}}\,, 
\end{align}
with $\mathcal{Z}^{(S)}(T_i)$ as the partition function in the overdamped case 
\begin{align}
    \mathcal{Z}^{(S)}(T_i) \equiv \int \dd x \, e^{- \frac{V(x)}{k_BT_i}}\,. 
\end{align}
Substituting $D_p(T) = m k_B T$ to the expression for the overlap in Eq.~\eqref{eq:a2-under} we get 
\begin{align}
\label{eq:a2-under-final}
a_2(T_i,T_b) =\left(1- \frac{1}{\epsilon^2}\frac{\mu_2}{\omega_0}\right) a_2^{(S)}(T_i,T_b) +
    \frac{1}{\epsilon^2}\frac{k_B\left(T_i - T_b
    \right)}{2 m \omega_0 ^2} \left\langle\left(u_2 ^{(S)}(x)\right)''\right\rangle_{f^{(S)}_\mathrm{eq}(T_i)} 
+\mathcal{O}\left(\epsilon^{-3}\right)\,.  
\end{align}
Recall that, $\mu _2 = \epsilon \lambda_2 ^{(S)}$, from Eq.~\eqref{eq:eig-Smo-eps}. 
Note that corrections that are odd in momentum do not contribute to $a_2$, since the initial momentum distribution is even in $p$.

Based on Eq.~\eqref{eq:a2-under-final}, we conclude that the leading term in the underdamped overlap coefficient is the overdamped overlap coefficient. Therefore, for sufficiently large damping, any robust overdamped Mpemba effect persists under the inclusion of inertia. The finite-damping corrections are of order \(\epsilon^{-2}\) and generally shift the location of the zero of \(a_2\). Thus, the strong-Mpemba temperature is perturbatively close to, but not exactly identical to, its overdamped value. Also, importantly, this derivation does not rely on the specific form of the potential and is therefore valid for an arbitrary confining potential.

\section{Numerical Methods} 
\subsection{Klein-Kramers equation}
\label{sec:KKnumericalMethods}
To characterize the Mpemba effect in the underdamped Klein-Kramers system, we employ a finite-difference discretization on a 
2D 
grid in phase space $(x, p)$. To ensure numerical stability and prevent unphysical negative probabilities, we utilize a Scharfetter-Gummel-type discretization scheme~\cite{chang1970practical}. This approach is essential for maintaining the Boltzmann distribution as the exact steady state of the system, even on relatively coarse grids. The numerical flux is constructed using the Bernoulli function, defined as:
\begin{equation}
    B(z) = \frac{z}{e^z - 1},
\end{equation}
which is implemented in our code to handle the exponential dependencies of the drift and diffusion terms. By incorporating this scheme, we effectively discretize the 
\ac{KK} 
operator $\mathcal{L}$ into a sparse matrix representation. This discretization also ensures that the Boltzmann distribution is the (unique) steady state of the discretized operator 
and respects the ``generalized'' detailed balance (appropriate momentum-reversal symmetry of the Klein--Kramers dynamics).

The eigenvalues and corresponding eigenvectors are computed using the Arnoldi iteration method, specifically targeting the modes with the 
largest real parts below zero to identify the dominant relaxation timescales. To construct the relaxation coefficients, we project the initial equilibrium distribution onto the relaxation modes. 

\begin{figure}[h]
    \centering
    \includegraphics[width=0.45\linewidth]{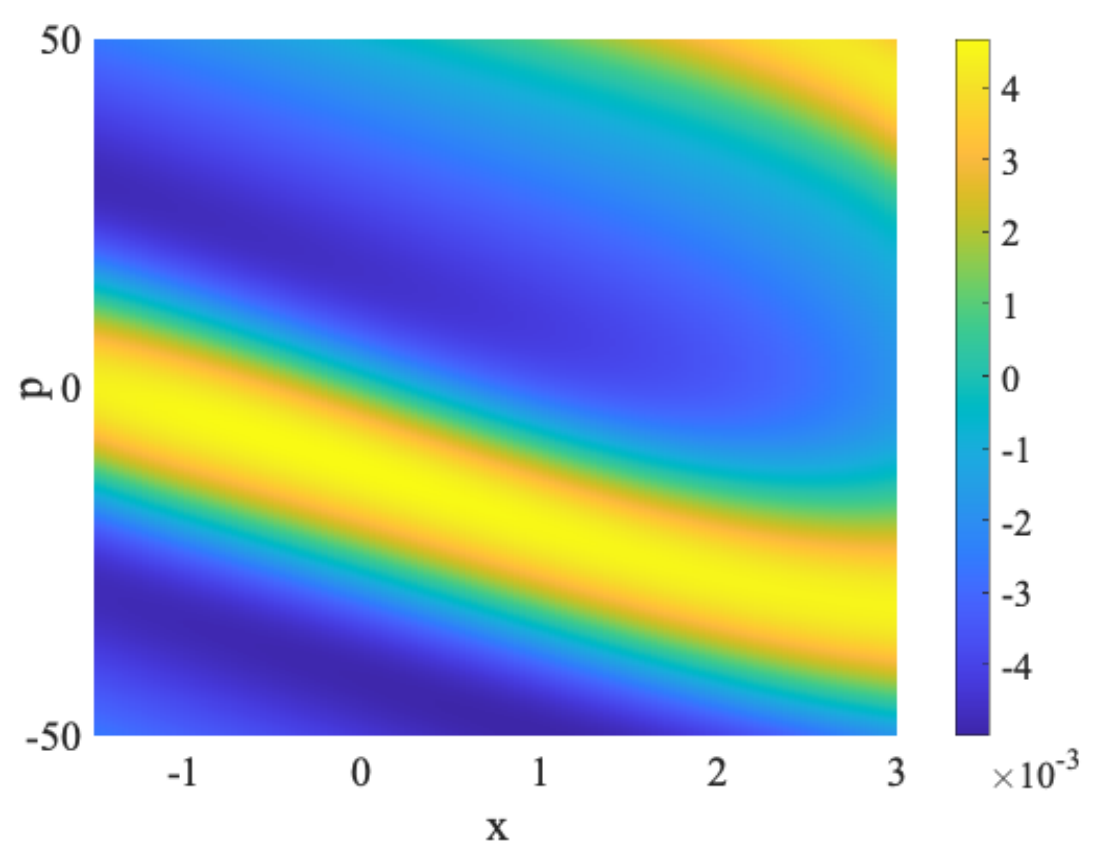}   \includegraphics[width=0.45\linewidth]{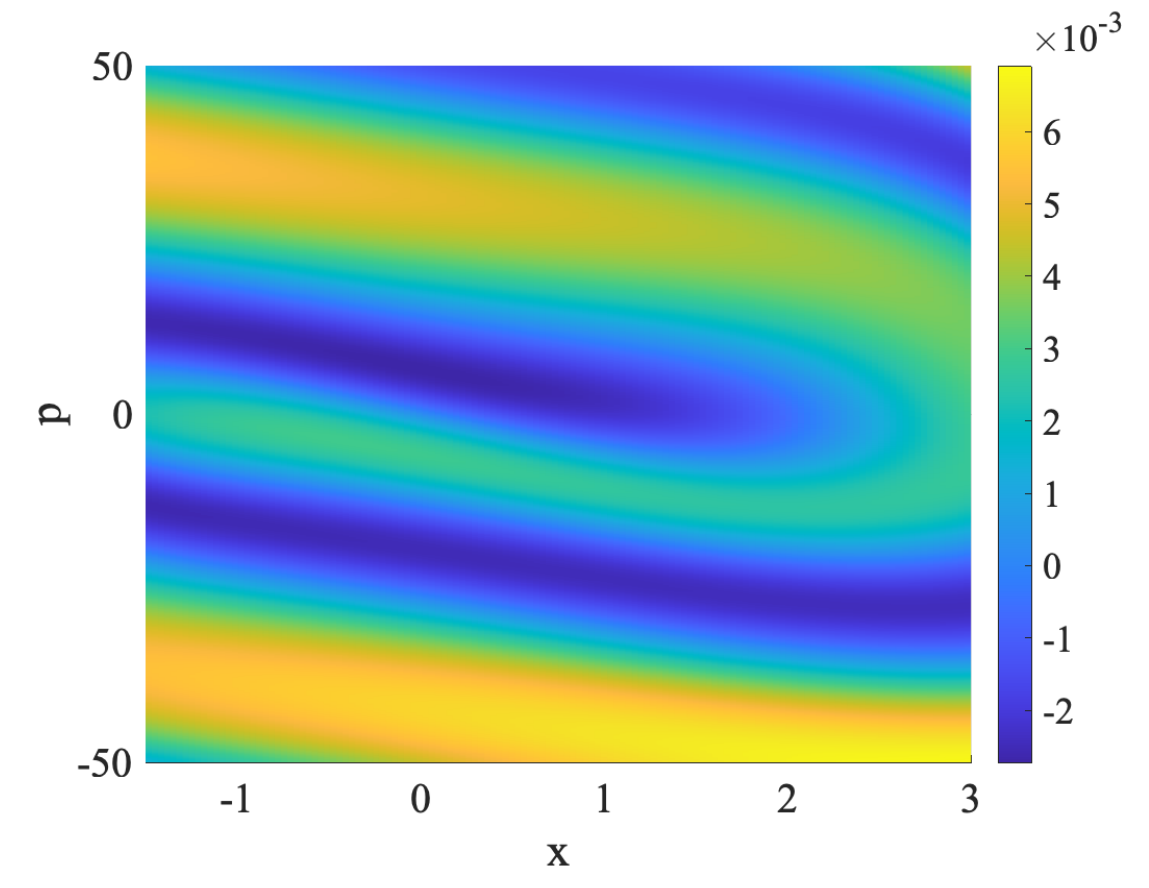}
    \caption{The left eigenfunction $u_2(x,p)$ for $T_b=14$ and at two values of dissipation $\gamma$. Left: For dissipation $\gamma = 8$ there is a strong inverse Mpemba effect; right: $\gamma=4$,  there is no Mpemba effect. The panels are calculated using the numerical method discussed in Sec.~\ref{sec:KKnumericalMethods}.}
    \label{fig:u2}
\end{figure}

In Fig.~(\ref{fig:u2}) we plot $u_2(x,p)$ for $T_b=14$ and two values of $\gamma$: $\gamma=8$ in the left panel and $\gamma=4$ in the right panel. These can provide some intuition for the existence and nonexistence of the inverse Mpemba effect: For very cold initial condition, 
the Boltzmann distribution $f_\mathrm{eq}(T_i)$ is concentrated in the $p$ direction at a small band around zero, and the width of this band changes with $T_i$.
In the $x$ direction, at very cold temperatures most of the probability is concentrated around the global minima, and some smaller fraction is near the other local minimum, where the ratio between these two is set by the temperature. For both values of $\gamma$, at low enough $T_i$ the projection $\langle u_2|f_\mathrm{eq}(T_i)\rangle$ essentially measures fractions of probability in the left vs. right well, much like the intuition in the overdamped case \cite{Lu2017Significance}. However, the width of the band in the $p$ direction for which this interpretation applies is very different for the two values of $\gamma$: It is thick for $\gamma=8$ and very thin for $\gamma=4$. Therefore, for $\gamma=8$ there is a $T_b$ for which the integration in this band is zero, whereas for $\gamma=4$ the band is too narrow and there is no effect. Similar intuition can be gained for other values.  

\subsection{Langevin dynamics} \label{sec:Langevin}


A numerical diagonalization of a large matrix, such as the \ac{KK} operator, is effective but time-consuming. Here, we use an efficient method to probe the existence of a strong \ac{ME} using the stochastic setting (See Langevin equation in Section~2 of the main text). Instead of scanning 
every set of initial and final temperatures, this method considers only the final one and two initial temperatures, defining an interval. The result is sensitive to an odd number of zero crossings of the coefficient of slow relaxation, in the given interval of initial temperatures. In addition, it serves to validate the results obtained by diagonalization.

We numerically solve the second-order differential equation by reduction of order, i.e., transforming it to a set of two first-order differential equations (e.g.~\cite[Chapter~4.2]{Amir_2020}). The resulting update formula is given by:
\begin{align}
    \left\{ \begin{array}{l}
         x(t+\Delta t) = x(t) + \frac{p(t)}{m} \Delta t \,,\\
         p(t+\Delta t) = p(t) - \leri{\partial_x V + \frac{\gamma}{m} p(t)} \Delta t 
         + \sqrt{2D} \eta(t) \sqrt{\Delta t} \,.
    \end{array} \right.
\end{align}
The potential was chosen to be similar to the one used in~\cite{Kumar2020Exponentially}, as presented in Fig.~1 (main text):
$V(x)= 2x^4 - 4x^2-0.65 x$, 
with an asymmetric spatial domain $x \in [-1.5, 3.5]$ and reflective boundary conditions.

In order to prove the existence of \ac{ME} in a stochastic setting, one may construct \ac{PDF} and compute its distance from the final state. However, constructing the full \ac{PDF}, especially in the underdamped regime, requires a large number of realizations. Moreover, one theoretically needs to scan every pair of initial and final temperatures. We therefore follow a different scheme as an indication of a strong \ac{ME}~\cite{gal2020precooling} (see Fig.~4 of the main text). For simplicity, we explain this method in a discrete setting, although the same approach could be used in a continuous one.
Consider the high-dimensional space of \ac{PDF}s, in which the equilibrium locus is a \ac{1D} curve. In a typical \ac{ME} experiment, the final steady state resides on the locus, at $f_\text{eq}(T_b)$. It is obvious that on that point, all coefficients besides $a_1 = 1$ vanish, as well as $a_2$. In fact, every point of the equilibrium locus is embedded in a hyperplane (of co-dimension one) spanned by the fast eigenvectors $\{\vec{v}_i\}$, $i \geq 3$, of the corresponding $T_b$. On that hyperplane, $a_2$ vanishes, and as a result, the relaxation to $f_\text{eq}(T_b)$ is exponentially faster. Moreover, this hyperplane divides the space of \ac{PDF}s into two regions, according to the sign of $a_2$: positive where the final stages of relaxation are along $\vec{v}_2$ and negative where they are against. For a strong \ac{ME} to take place, this hyperplane must cross the equilibrium locus in at least another point, $T_\text{SME}$, in addition to $T_b$. Assuming $a_2$ crosses zero continuously, at initial temperatures slightly above and slightly below $T_\text{SME}$, the final stages of relaxation towards $f_\text{eq}(T_b)$ will have opposite directions. Thus, an approach at the late stages of relaxation from opposite directions of two different initial temperatures, while all other parameters are identical, indicates a strong \ac{ME} at some intermediate temperature. This indication provides evidence for a strong \ac{ME} even when projected onto a lower subspace. As a result, the number of required realizations can be decreased substantially.

\bibliography{references}